\documentclass[journal]{IEEEtran}
\usepackage{amssymb}
\usepackage[figuresright]{rotating}
\usepackage[tableposition=top]{caption}
\usepackage{amsmath,subfigure,epsfig,bbding}
\usepackage{epstopdf}
\usepackage{multicol}
\usepackage{url}
\usepackage{booktabs}
\usepackage{multirow}
\usepackage{threeparttable}
\usepackage{color}
\usepackage[mathscr]{eucal}

\DeclareCaptionFont{blue}{\color{blue}}
\usepackage{makecell}
\usepackage{amsmath}
\usepackage[linesnumbered,ruled,vlined]{algorithm2e}
\usepackage{algpseudocode}
\usepackage{amsmath}
\usepackage[colorlinks=true,
            linkcolor=black, 
            anchorcolor=black, 
            citecolor=black, 
            urlcolor=blue
            ]{hyperref}

\def\hlinew#1{%
  \noalign{\ifnum0=`}\fi\hrule \@height #1 \futurelet
   \reserved@a\@xhline}
\newcommand{\PreserveBackslash}[1]{\let\temp=\\#1\let\\=\temp}
\newcolumntype{C}[1]{>{\PreserveBackslash\centering}p{#1}}
\newcolumntype{R}[1]{>{\PreserveBackslash\raggedleft}p{#1}}
\newcolumntype{L}[1]{>{\PreserveBackslash\raggedright}p{#1}}

\begin{document}

\title{HVS Revisited: A Comprehensive Video Quality Assessment Framework}

\author{Ao-Xiang~Zhang,~\IEEEmembership{Student Member,~IEEE,} Yuan-Gen~Wang,~\IEEEmembership{Senior Member,~IEEE}\\
Weixuan Tang,~\IEEEmembership{Member,~IEEE},
Leida Li,~\IEEEmembership{Member,~IEEE}, Sam Kwong,~\IEEEmembership{Fellow,~IEEE}

\thanks{

A.-X. Zhang and Y.-G. Wang are with the School of Computer Science and Cyber Engineering, Guangzhou University, Guangzhou 510006, China (e-mail: zax@e.gzhu.edu.cn; wangyg@gzhu.edu.cn).

W. Tang is with the Institute of Artificial Intelligence and Blockchain, Guangzhou University, Guangzhou 510006, China (email: tweix@gzhu.edu.cn).

L. Li is with the School of Artificial Intelligence, Xidian University, Xian 710071, China (e-mail: ldli@xidian.edu.cn).

S. Kwong is with the Department of Computer Science, City University of Hong Kong, Hong Kong (e-mail: cssamk@cityu.edu.hk).
}}

\maketitle

\begin{abstract}
Video quality is a primary concern for video service providers.
In recent years, the techniques of video quality assessment (VQA) based on deep convolutional neural network (CNN) have been developed rapidly.
Although existing works attempt to introduce the knowledge of the human visual system (HVS) into VQA, there still exhibit limitations that prevent full exploitation of HVS, including an incomplete model by few characteristics and insufficient connections among these characteristics.
To overcome these limitations, this paper revisits HVS with five representative characteristics, and further reorganizes their connections.
Based on the revisited HVS, a no-reference VQA framework called HVS-5M (NRVQA framework with five modules simulating HVS with five characteristics) is proposed.
It works in a domain-fusion design paradigm with advanced network structures.
On the side of the spatial domain, the visual saliency module applies SAMNet to obtain a saliency map.
And then, the content-dependency and the edge masking modules respectively utilize ConvNeXt to extract the spatial features, which have been attentively weighted by the saliency map
for the purpose of highlighting those regions that human beings may be interested in.
On the other side of the temporal domain,
to supplement the static spatial features,
the motion perception module utilizes SlowFast to obtain the dynamic temporal features.
Besides, the temporal hysteresis module applies TempHyst to simulate the memory mechanism of human beings, and comprehensively evaluates the quality score according to the fusion features from the spatial and temporal domains.
Extensive experiments show that our HVS-5M outperforms the state-of-the-art VQA methods.
Ablation studies are further conducted to verify the effectiveness of each module towards the proposed framework.
\end{abstract}


\section{Introduction}


\IEEEPARstart{R}{ecent} years have witnessed an explosive growth of
``we-media''.
It is estimated that there are about 4 billion video views per day on Facebook \cite{Facebook}. However, storing and delivering these vast amounts of video data
greatly stresses video service providers \cite{Mao2022}.
It is necessary to apply video coding to
reduce the storage capacity, and balance the tradeoff between the coding efficiency and the video quality.
Therefore, video quality assessment (VQA) has become a hot research topic \cite{Zhang2019}, \cite{Wu2021}, \cite{CSPT}, \cite{Xing2022}.
Subjective VQA is a manual rating by human beings which is time-consuming and labor-consuming \cite{Ou2021}.
By contrast, objective VQA is an automatic predicting by machines, and thus is more widely used in real application scenarios.
Since the scoring indicator of VQA, i.e., mean opinion score (MOS), is related to the visual effect of human beings, it is of great benefit to introduce human visual system (HVS) into VQA \cite{Xian2022,Zhang2022}.

Early HVS-based VQA methods utilized hand-crafted features to handle synthetic distortions.
In order to accurately simulate the texture masking of HVS, Ma \emph{et al.} \cite{Ma2013} developed a mutual masking strategy to extract the spatial information of video. Galkandage \emph{et al.} \cite{Galkandage2017} incorporated binocular suppression and recurrent excitation.
Saad \emph{et al.} \cite{VBLIINDS} proposed a non-distortion specific evaluation model relied on the video scenes in discrete cosine transform domain, and analyzed the types of motion that occurred in the video to predict the video quality.
Korhonen proposed TLVQM \cite{TLVQM} to reduce the complexity of feature extraction with a two-level method, which obtained the low and high complexity features from an entire video sequence and several representative video frames, respectively.
Wang and Li \cite{Wang2007} proposed a statistical model of human visual speed perception, which can estimate the motion information and the perceptual uncertainty in the video.




Recently, with the rapid development of deep learning techniques, CNN-based (Convolutional Neural Network)
VQA methods have significantly improved the evaluation of in-the-wild videos.
The in-the-wild videos are referred to as the authentically distorted ones, which are often hard to be annotated due to the lack of pristine ones.
To solve the problem of insufficient training data,
You and Korhonen \cite{You2019} utilized the extended short-term memory network to predict the video quality based on the features extracted from small video cube clips by 3D-CNNs. Li \emph{et al.} \cite{VSFA} applied the gated recurrent unit (GRU) \cite{GRU} to obtain the video quality score according to the frame-level content features extracted from ResNet \cite{ResNet}. Zhang and Wang \cite{TiVQA} utilized texture features to complement content features, further improving \cite{VSFA}.
However, all these methods had not fully exploited the temporal information within videos, which led to poor performance on LIVE-VQC containing rich motion-related contents.
Chen \emph{et al.} \cite{RIRNet} proposed to fuse motion information from different temporal frequencies in an efficient manner, and further applied a hierarchical distortion description to quantify the temporal motion effect.
Li \emph{et al.} \cite{Li2022} proposed a model-based transfer learning approach and applied motion perception to VQA.
Besides, some methods attempted to introduce visual saliency into VQA. For instance, Guan \emph{et al.} \cite{Guan2022} established a quality-aware visual attention module to obtain the frame-level quality scores, which were integrated into the video quality score through an end-to-end structure of visual and memory attention.
Vagar \cite{Vagar2022} proposed a parallel CNN structure, which firstly extracted the temporally pooled and saliency weighted features of the video, and then independently mapped them to the quality scores for further fusion.

\begin{figure}[t]\centering
\setlength{\belowcaptionskip}{-0.18cm}
{\includegraphics[width=3.85in]{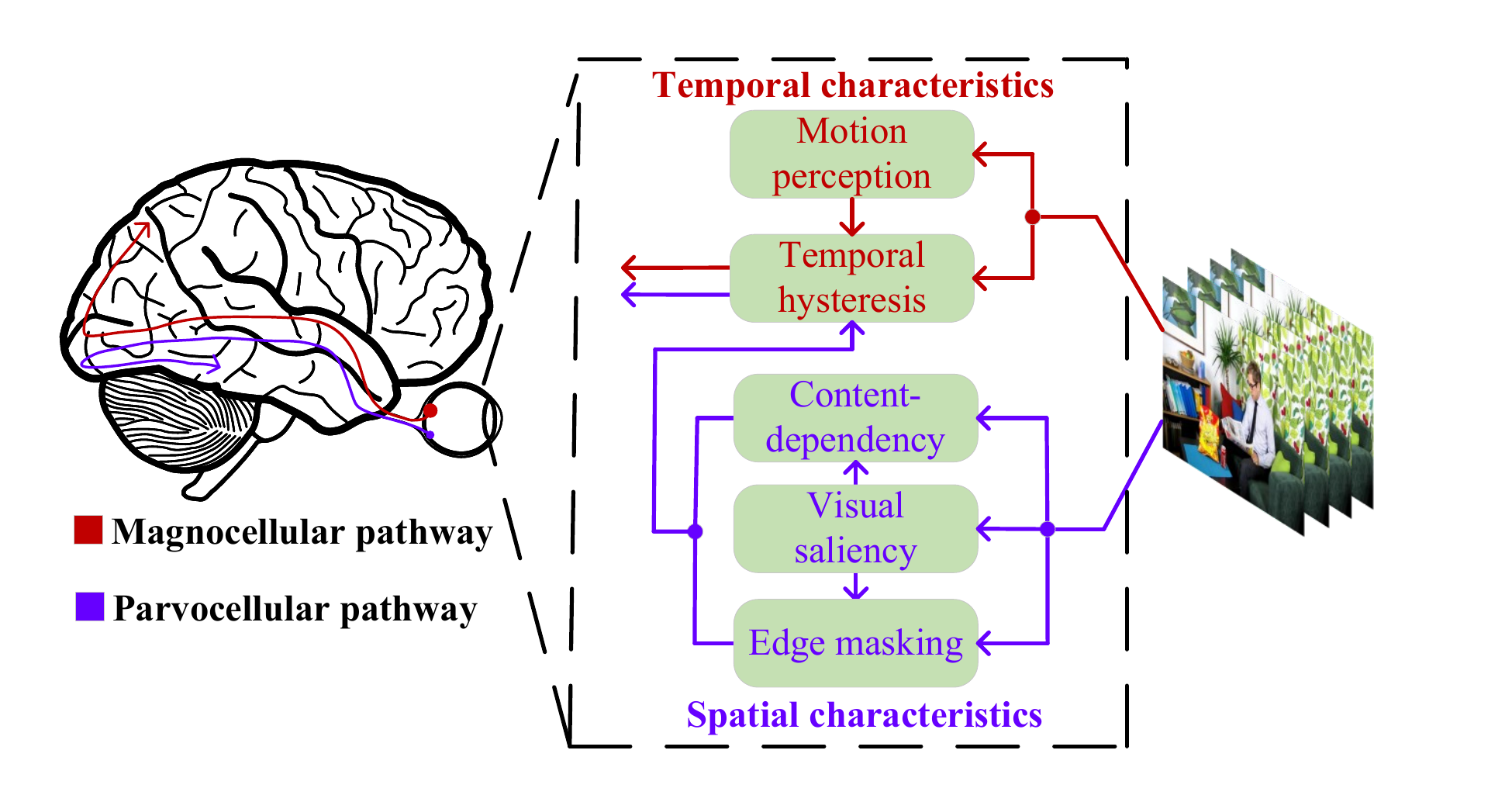}}

\caption{Illustration of our revisited version of HVS.}
\label{fig:HVS}
\end{figure}

Despite their good performance, these methods have some potential limitations.
Firstly, only few characteristics of HVS have been exploited, and thus the simulated sensory function is incomplete. Besides, the connections among different characteristics have not been well organized, failing to facilitate their applications to VQA.
Secondly, some methods have not fully considered temporal information. The combination of spatial and temporal information is to be explored.
Thirdly, the effect of edge masking has not been introduced in VQA. In fact, shape edges can not only be utilized to conceal distortions in edge masking, but also be served as effective spatial features.
Fourthly, although early attempts have been made, how to effectively apply visual saliency to VQA is still challenging.

To address the above problems, this paper proposes a
general no-reference video quality assessment framework called HVS-5M (NRVQA framework with five modules simulating HVS with five characteristics).
The foundation of HVS-5M is our revisited version of HVS, wherein the connections of the five representative characteristics of HVS are reorganized.
On this basis, HVS-5M is designed in the domain-fusion paradigm.
Specifically, on the side of the spatial domain, the edge masking and the content-dependency modules are utilized to extract the frame-level spatial features, which are then weighted by the saliency map from the visual saliency module according to the attention mechanism.
On the other side of the temporal domain, to supplement the spatial features,
the motion perception module is utilized to extract the video-level temporal features.
Furthermore, the temporal hysteresis module simulates the memory mechanism of human beings, and outputs the quality score according to the fusion features integrated from the spatial static ones and temporal dynamic ones.
By this means, the quality of a given video can be comprehensively represented from different aspects.

The contributions of this work are summarized as follows.
\begin{itemize}
\item The mechanism of HVS is revisited, wherein five representative characteristics are selected to model the function of sensory organ in a relatively simple and comprehensive manner, and their connections are reorganized to facilitate its application to VQA.

\item 
    A video quality assessment framework simulating HVS with five modules, called HVS-5M, is proposed, wherein these modules cooperatively work in the domain-fusion paradigm.
    In particular, to the best of our knowledge, this is the first to introduce edge masking and a new scheme to apply visual saliency to VQA.

\item Experimental results show that our proposed HVS-5M achieves state-of-the-art (SOTA) performance on various mainstream video datasets, including four in-the-wild ones. Ablation studies are further conducted to verify the effectiveness of its different modules.
\end{itemize}

The rest of paper is organized as follows. HVS is first revisited from a neurophysiology perspective in Section \ref{HVS revisited}.
Then, the proposed HVS-5M is described in Section \ref{proposed method}.
Extensive experimental results are presented in Section \ref{sec:exp}.
Finally, conclusions are drawn in Section \ref{conclusion}.

\section{Human Visual System Revisited} \label{HVS revisited}

Human visual system (HVS) is responsible for detecting and interpreting the perceived spectral information to build a representation of the surrounding environment, which consists of sensory organ and parts of the central nervous system.

Specifically, the function of sensory organ has many complicated characteristics.
In order to demonstrate the function of sensory organ in a relative simple but comprehensive manner,
we revisit HVS, and formulate it with five representative characteristics.
\textit{Visual saliency} is a bottom-up, stimulus-driven signal, which indicates that a specific location is sufficiently different from the surrounding environment and deserves human attention \cite{Visual Saliency1}. \textit{Content-dependency} is referred to as a phenomenon that human preference is highly dependent on the observed content \cite{content-dependency4}.
\textit{Edge masking} indicates that the effect of masking is more likely to occur at positions with richer edge information \cite{Edge Masking}.
\textit{Motion perception} is the process of inferring the speed and direction of various elements in a dynamic scene.
\textit{Temporal hysteresis} \cite{Temporal-Memory} indicates that
the memory of elements with bad impression can last for longer than those with good impression.


\begin{figure*}[t]\centering
\setlength{\belowcaptionskip}{-0.18cm}
{\includegraphics[width=7.25in]{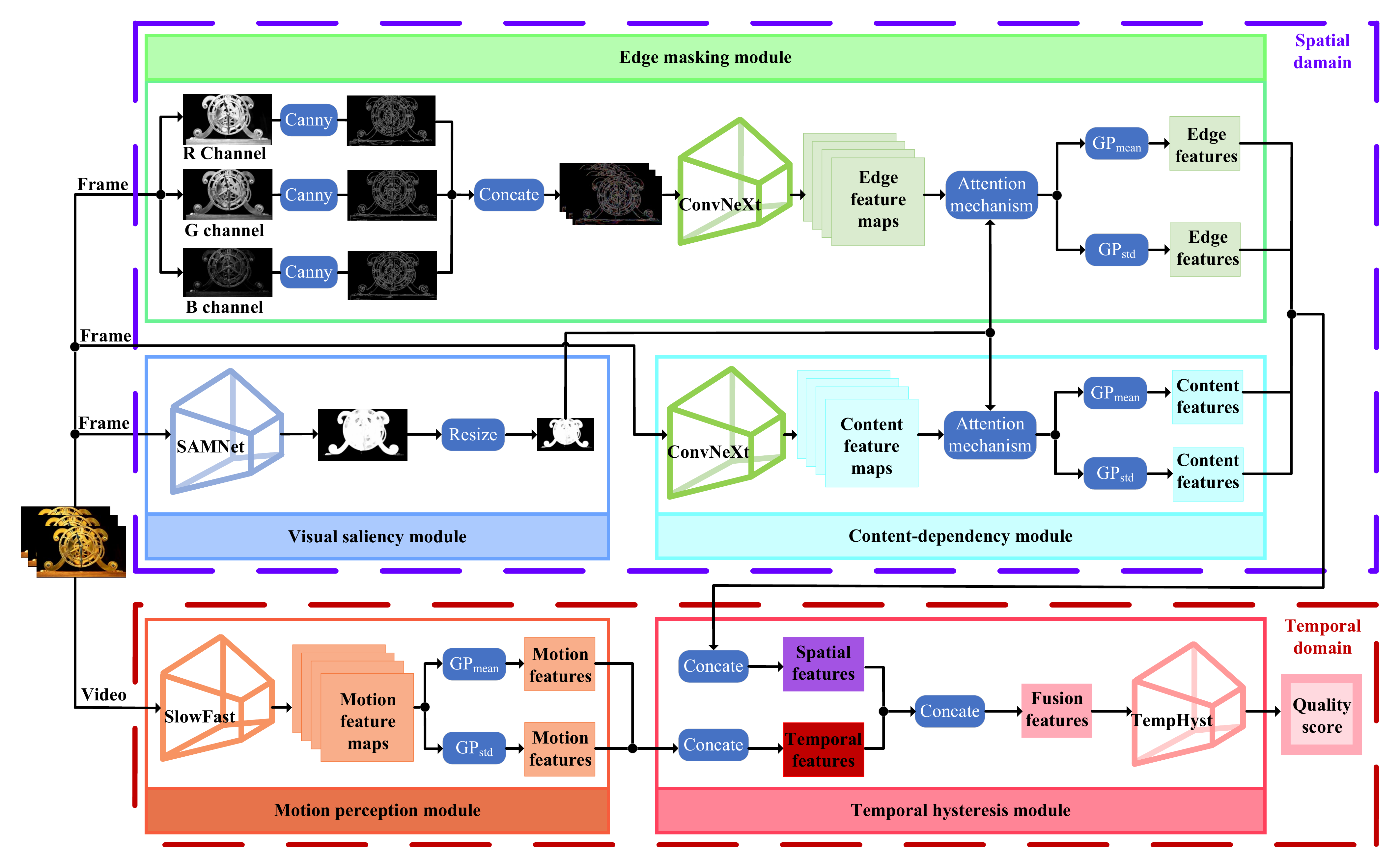}}
\caption{Illustration of the proposed HVS-5M, which includes five modules.}\label{framework}
\end{figure*}

On this basis, in order to apply HVS to VQA in a more efficient manner, we try to establish connections for the five characteristics in HVS, as shown in Fig. \ref{fig:HVS}.
According to the existing study \cite{P&M-path}, the parvocellular pathway (P-path) and magnocellular pathway (M-path) are two major pathways of the central nervous system.
Specifically, P-path can distinguish subtle spatial details \cite{P-cells}, while M-path is capable of detecting temporal motion information \cite{M-cells1}.
Therefore, we correspondingly divide those five characteristics into spatial or temporal ones.
On the side of the spatial characteristics, human beings could be attracted by salient regions (reflected in visual saliency) at first sight.
And then, under the guidance of the salient regions,
content (reflected in content-dependency) and edge (reflected in edge masking) are used to represent semantics and details, respectively.
On the other side of the temporal characteristics, motion information (reflected in motion perception)
is used to represent dynamic temporal changes.
Finally, the above semantics, details, and temporal changes are fused into a sequence of elements along time flow,
wherein the elements with bad impression are highlighted (reflected in temporal hysteresis).
By this means, these five characteristics can collaboratively work in HVS.
In this paper, we focus on reorganizing the connections of the five representative characteristics within our revisited version of HVS, and take it as a starting point for the research of VQA.



\section{Proposed Method} \label{proposed method}

In order to simulate the HVS mechanism, the proposed HVS-5M follows the domain-fusion paradigm and consists of five modules from two domains, as shown in Fig. \ref{framework}.

The first branch operates on frame level in spatial domain.
The visual saliency module extracts the saliency map for each frame.
Then, the content-dependency module and the edge masking module are responsible for extracting the content and the edge feature maps, respectively. These two feature maps are further adjusted by the saliency map in the attention manner for the purpose of highlighting the key regions, and then integrated into statistics as spatial features.
The second branch operates on video level in temporal domain, wherein the motion perception module aims at capturing the motion feature maps of a video sequence, and then combines them into temporal features.
Finally, the temporal hysteresis module simulates the memory mechanism of human beings,
and comprehensively evaluates the quality score according to the fusion features
from the spatial and temporal domains.
The detailed descriptions of each module are given in the following subsections, and the notations of features are given in Table \ref{variable}.

\subsection{Visual Saliency Module}

In HVS, it is widely acknowledged that human attention is attracted by visual salient region within an image \cite{Visual Saliency1,Visual Saliency2}.
To simulate HVS, the proposed HVS-5M also adopts the saliency map as an attention mask to highlight those regions that human beings may be interested in.
In the following of this paper, the video is supposed to have $N$ frames and the $n$-th frame is denoted as $\mathbf{I}_n$.
Considering both accuracy and computational complexity,
a lightweight network SAMNet \cite{SAMNet} pre-trained on ImageNet-22k \cite{ImageNet} is adopted to extract the saliency map $\mathbf{A}_n$ for each frame as
\begin{equation} \label{Saliency Detection}
\mathbf{A}_{n}=\text{SAMNet}\left( \mathbf{I}_{n} \right) \ .
\end{equation}
In SAMNet, a stereoscopically attentive multi-scale module is designed for effective and efficient multi-scale learning,
which enables each channel at each spatial position to adjust the weights of each branch.
Based on this module, a lightweight encoder-decoder network is utilized for salient object detection.
Note that ImageNet-22k is a large-scale image dataset with 22,000 categories.
The SAMNet pre-trained on ImageNet-22k has the ability to extract saliency map for a variety of images, and can fulfill the demand of visual saliency in HVS.
The output of SAMNet is the saliency map $\mathbf{A}_n$.

\begin{figure}[t]\centering

\setlength{\belowcaptionskip}{-0.18cm}
\centerline{\includegraphics[width=3.75in]{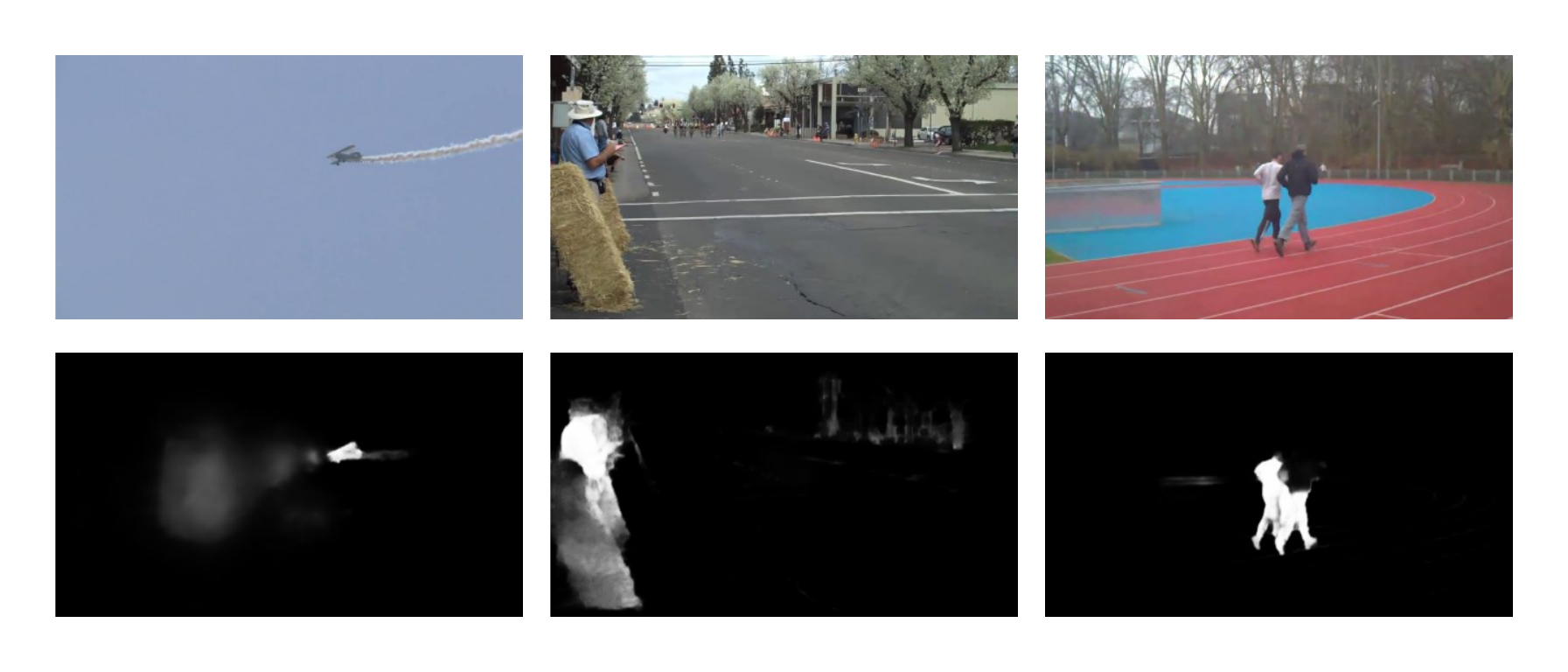}}
\setlength{\abovecaptionskip}{-0.15cm}

\caption{The video frame and its corresponding saliency map.} \label{Saliency map}
\end{figure}

Note that each element of $\mathbf{A}_n$ can be interpreted as probability, which lies within the range of [0,1].
Via element-wise multiplying such saliency map with the feature maps extracted in the subsequent modules, it is expected that the model can
suppress the features within regions that human beings are not interested in.
However, as shown in Fig. \ref{Saliency map}, it can be observed that within the saliency map extracted by SAMNet, a majority of elements tend to 0.
Under the guidance of such saliency map, the model may excessively suppress some regions which may still contain certain useful spatial information, and cannot
bring the potential of visual saliency to full play.
Therefore, to better adapt the saliency map to the task of VQA, we perform adaptive adjustment on the saliency map, which is formulated in Eq. \eqref{eq:adaptive}.
Specifically, we multiply the elements in $\mathbf{A}_n$ by 255.
To further distinguish the significant regions from the non-significant ones, the elements greater than threshold, denoted as $h$, are added by 350, while those less than $h$ are added by 250.
And then, to make use of the saliency map in the subsequent modules, such saliency map is downsampled into the same size of the outputted feature maps in the content-dependency module and the edge masking module.
In our HVS-5M, $h$ is empirically set to 100.
\begin{equation}\label{eq:adaptive}
\left\{ \begin{array}{l}
	\widehat{\mathbf{A}}_n=\text{resize}\left( \mathbf{A}_n \times 255 +250 \right), \quad \text{if} \quad \mathbf{A}_n < h,
\\[5pt]
	\widehat{\mathbf{A}}_n=\text{resize}\left( \mathbf{A}_n \times 255 +350 \right),  \quad \text{otherwise}.\\
\end{array} \right.
\end{equation}




\begin{table}[t]
\caption{The implication and dimension of each variable.}\label{variable}
\centering
\fontsize{7}{7}\selectfont
\renewcommand{\arraystretch}{1.15}
\setlength{\tabcolsep}{0.1mm}{
\begin{tabular}{ccc}
\toprule[1.1pt]
Notation & Implication                    & Dimension          \\ \midrule
$\mathbf{I}_n$       & Video frame           & $\left( H, W, 3 \right)$            \\
$\mathbf{A}_n$       & Saliency map        & $\left( H, W, 1 \right)$             \\
$\widehat{\mathbf{A}}_n$      & Resized saliency map  & $\left( H/32, W/32, 1 \right)$        \\
$\mathbf{C}_n$       & Content feature maps           & $\left(  H/32, W/32, 2048 \right)$ \\
$\widetilde{\mathbf{C}}_{n}$       & Content feature maps with attention           & $\left( H/32, W/32, 2048 \right)$ \\
${\mathbf{C}^\text{mean}_{n}}$       & Mean of content feature maps         & $\left( 2048 \right)$ \\
 ${\mathbf{C}^\text{std}_{n}}$       & Std of content feature maps           & $\left(  2048 \right)$ \\
$\mathbf{D}_n$      & Edge maps          & $\left( H, W, 3 \right)$ \\
$\mathbf{E}_n$       & Edge feature maps           & $\left(  H/32, W/32, 2048 \right)$ \\
${\widetilde{\mathbf{E}}_n}$       & Edge feature maps with attention          & $\left(  H/32, W/32, 2048 \right)$ \\
${\mathbf{E}^\text{mean}_{n}}$       & Mean of edge feature maps                  & $\left( 2048 \right)$ \\
${\mathbf{E}^\text{std}_{n}}$       & Std of edge feature maps                          & $\left(  2048 \right)$ \\
$\mathbf{M}$      & Motion feature maps           & $\left( N/2, H/32, W/32, 256 \right)$ \\
${\mathbf{M}}^\text{mean}$      & Mean of motion feature maps                           & $\left(  N/2, 256 \right)$ \\
${\mathbf{M}}^\text{std}$      & Std of motion feature maps                           & $\left(  N/2, 256 \right)$ \\
$\mathbf{S}_{n}$       & Spatial features            & $\left(  8192 \right)$ \\
$\mathbf{T}$      & Temporal features        & $\left(  N/2, 512 \right)$ \\
$\left\{ \mathbf{S}_n \right\}_{n=1}^{N}$     & Spatial features of different frames & $\left(  N, 8192 \right)$ \\
$\text{Sample}\left( \left\{ \mathbf{S}_n \right\}_{n=1}^{N} \right)$      & Sampled spatial features        & $\left(  N/2, 8192 \right)$ \\
$\mathbf{F}_n$      & Fusion features         & $\left( N/2, 8704 \right)$ \\

 \bottomrule[1.1pt]
\end{tabular}}
\end{table}


\subsection{Content-Dependency Module}

Video frames contain a wealth of contents affecting the quality score of videos.
Such argument is supported by many subjective experiments \cite{content-dependency1,content-dependency2,content-dependency3, content-dependency4}, wherein images or videos with different contents but under the same condition may correspond to different subjective quality scores in HVS.
Therefore, the proposed HVS-5M utilizes the content-dependency module to capture the content information.
Firstly, ConvNeXt \cite{ConvNeXt} pre-trained on ImageNet-22k \cite{ImageNet} is utilized to extract the content feature maps $\mathbf{C}_n$ for each frame as
\begin{equation}\label{Content extract}
\mathbf{C}_n=\text{CNN}\left( \mathbf{I}_{n} \right).\\
\end{equation}
Inheriting some key components of Vision Transformer \cite{Vision Transformer}, ConvNeXt makes incremental changes to a standard ResNet \cite{ResNet}.
Such neural network is purely based on convolution and can take advantage of inherent inductive bias such as locality and translation variance.
ConvNeXt pre-trained on ImageNet-22k has the ability to distinguish different contents, and thus its intermediate features can effectively represent the contents of each frame within video.
Specifically, the feature maps in the `res5c' layer are adopted as the content feature maps $\mathbf{C}_n$.

Secondly, to highlight those regions that human beings are interested in, we apply attention mechanism on the extracted content feature maps $\mathbf{C}_n$ via the saliency map $\widehat{\mathbf{A}}_n$ from the visual saliency module as
\begin{equation}\label{}
\widetilde{\mathbf{C}}_{n}=\mathbf{C}_{n}\otimes \widehat{\mathbf{A}}_n,
\end{equation}
where $\otimes$ represents the element-wise multiplication for each channel of $\mathbf{C}_{n}$.
Thirdly, to further obtain the high-level statistics, we apply the global average pooling ($\text{GP}_\text{mean}$) and global standard deviation pooling ($\text{GP}_\text{std}$) on $\widetilde{\mathbf{C}}_{n}$ as
\begin{equation}\label{mean&std_for_content1}
\mathbf{C}^\text{mean}_{n}=\text{GP}_{\text{mean}}\left( \widetilde{\mathbf{C}}_{n} \right),
\end{equation}
\begin{equation}\label{mean&std_for_content2}
\mathbf{C}^\text{std}_{n}=\text{GP}_{\text{std}}\left( \widetilde{\mathbf{C}}_{n} \right).\\
\end{equation}
With these two statistics, the video frame can be relatively comprehensively described.



\subsection{Edge Masking Module}

\begin{figure}[tp]\centering
\setlength{\belowcaptionskip}{-0.18cm}
\subfigure[KoNViD-1k]
{\includegraphics[width=1.7in]{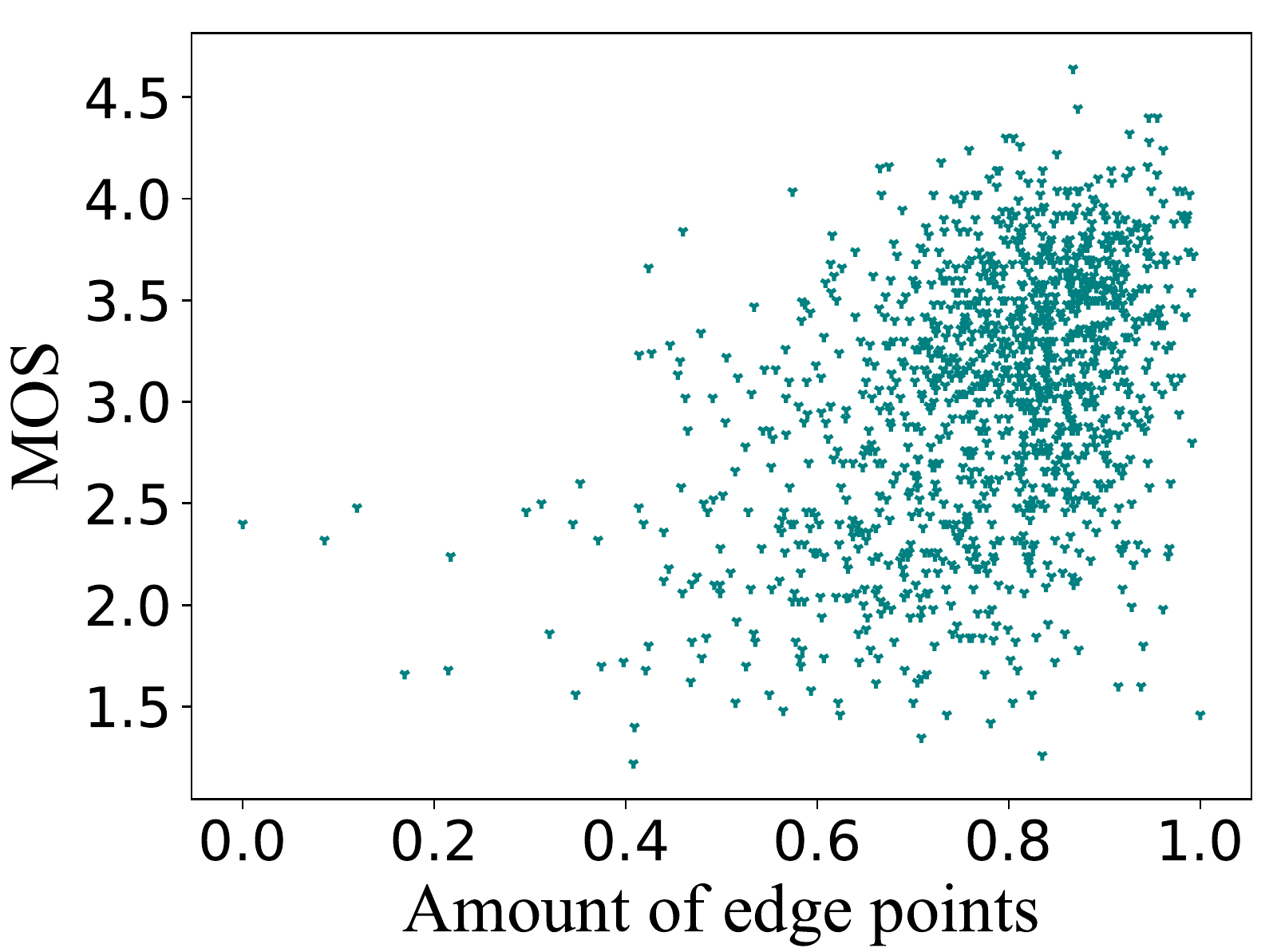}}
\subfigure[YouTube-UGC]
{\includegraphics[width=1.7in]{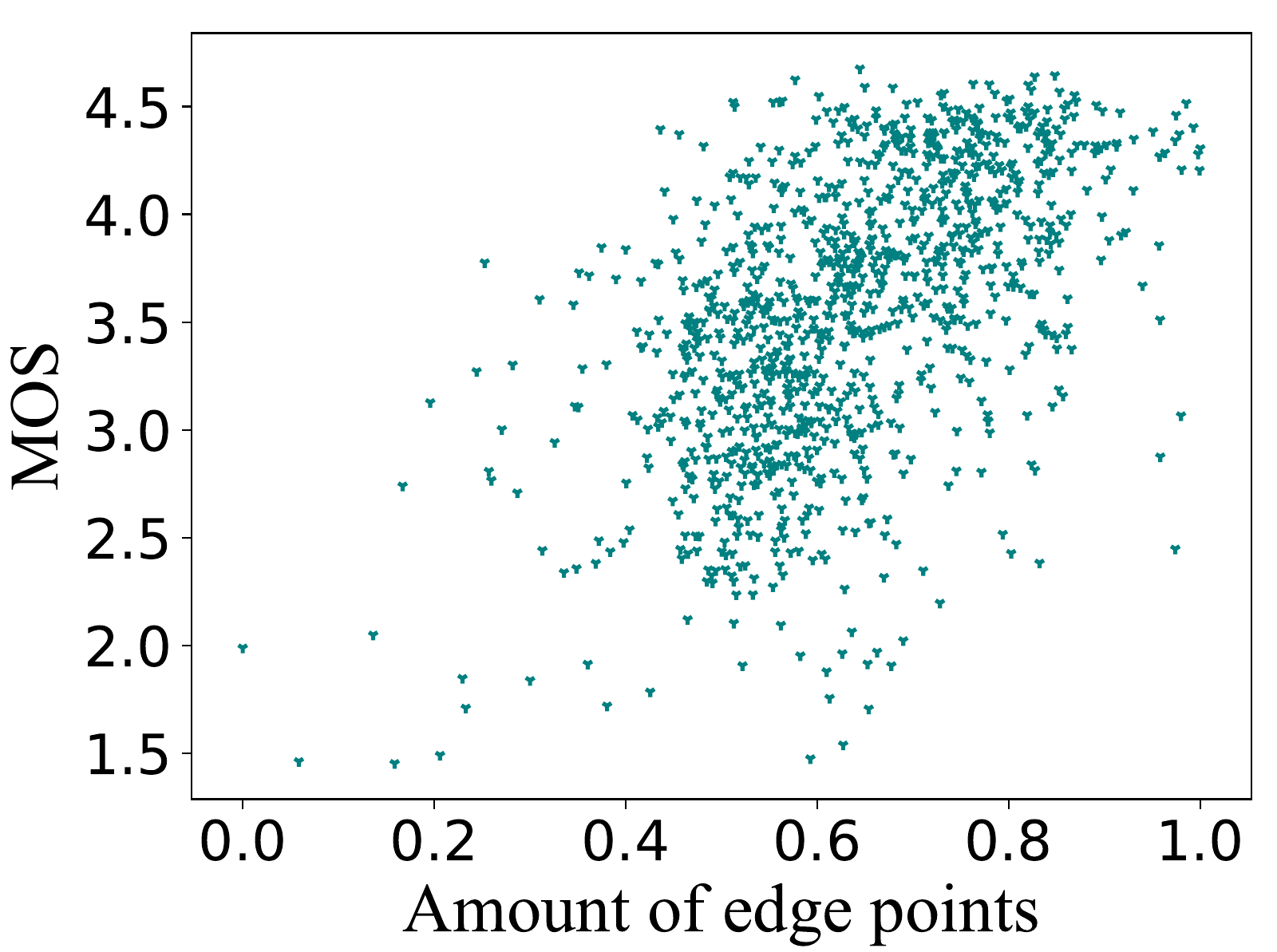}}
\subfigure[LIVE-VQC]
{\includegraphics[width=1.7in]{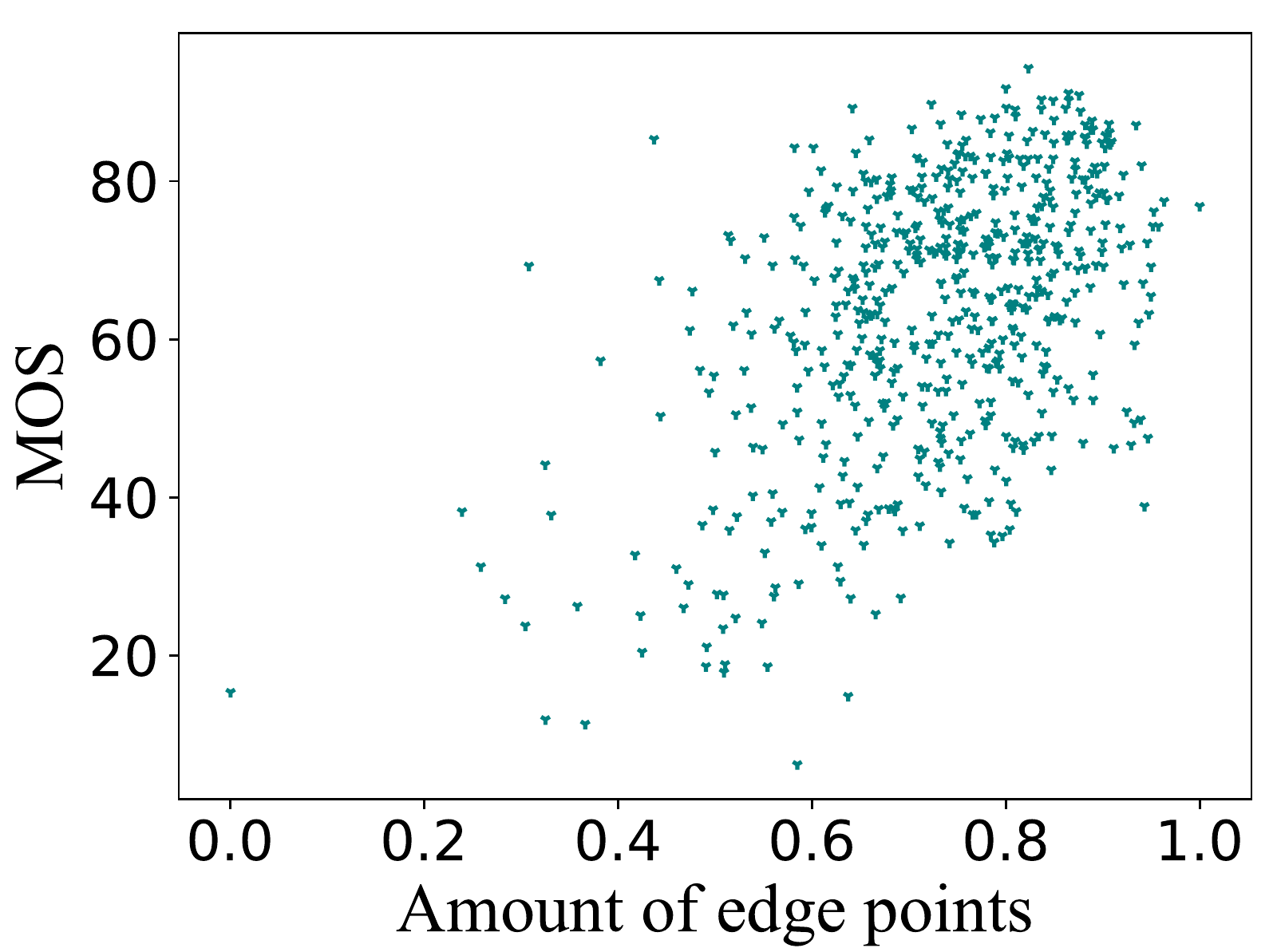}}
\subfigure[CVD2014]
{\includegraphics[width=1.7in]{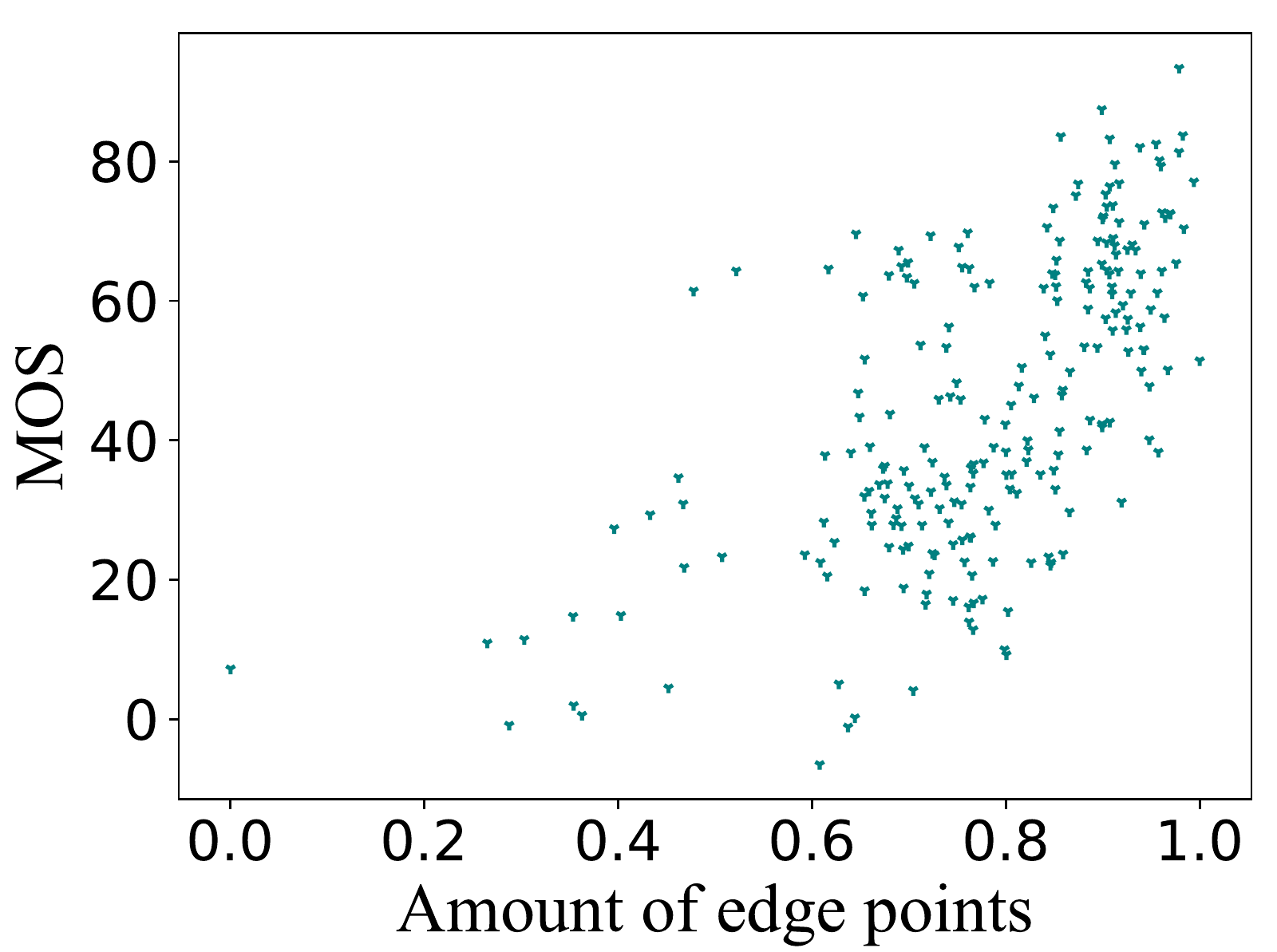}}

\vspace{-0.2cm}
\caption{The relationship between the amount of edge points in a frame and averaged within a video, and the MOS assigned for the video.
The edge points are calculated by Canny, and
the amount of edge points is normalized within [0,1].}
\label{Edge Point}
\end{figure}

In HVS, edge masking is referred to as the phenomenon that distortions can be concealed by sharp edges \cite{Edge Masking}.
Videos with richer edges can better hide distortions, and thus may have higher quality scores.
From Fig. \ref{Edge Point}, it can be observed that on four different datasets, the MOS is positively correlated with the amount of edge points.
Therefore, edge information has great impact on the performance of VQA.
However, merely supervised by MOS, it is rather difficult for existing models to capture proper edge information to simulate edge masking. In the proposed HVS-5M, edge masking module is utilized to explicitly extract plentiful edges based on Canny operator.

We divide each frame $\mathbf{I}_{n}$ into red, green, and blue channel as $\mathbf{I}_{n}^{\text{R}}$, $\mathbf{I}_{n}^{\text{G}}$, and $\mathbf{I}_{n}^{\text{B}}$.
These three channels are respectively processed to get the channel-wise edge map. The main steps are described as follows.
Firstly, Canny operator is utilized to calculate the gradient intensity of each element.
Then, the gradient intensity is compared with the upper threshold denoted as $u$ and the lower threshold denoted as $l$ to obtain the binary edge map.
Within the binary edge map, those pixels with gradient intensity larger than $u$, and part of those pixels with gradient intensity between $l$ and $u$, are considered as edge points and assigned with 255. The rest pixels are assigned with 0.
In our HVS-5M, $u$ and $l$ are empirically set to 140 and 5, respectively.
To this end, three channel-wise edge maps are concatenated into final edge maps $\mathbf{D}_n$ as
\begin{equation}
\mathbf{D}_{n}=\text{Concat} \left(  \text{Canny}_{u,l}\left( \mathbf{I}_{n\!\:}^{\text{R}} \right), \text{Canny}_{u,l}\left( \mathbf{I}_{n\!\:}^{\text{G}} \right), \text{Canny}_{u,l}\left( \mathbf{I}_{n\!\:}^{\text{B}} \right)\right).
\end{equation}

Afterwards, similar to those operations in the content-dependency module,
ConvNeXt \cite{ConvNeXt} pre-trained on ImageNet-22k \cite{ImageNet} is utilized to extract the edge feature maps $\mathbf{E}_n$ based on $\mathbf{D}_{n}$ as
\begin{equation}\label{}
\mathbf{E}_n=\text{CNN}\left( \mathbf{D}_{n} \right).\\
\end{equation}
And then, the saliency map $\widehat{\mathbf{A}}_n$ is incorporated in the attention manner as
\begin{equation}\label{}
{\widetilde{\mathbf{E}}_n}=\mathbf{E}_n \otimes \widehat{\mathbf{A}}_n,\\
\end{equation}
Finally, the statistics of mean and standard deviation are obtained as
\begin{equation}\label{}
{\mathbf{E}^\text{mean}_n}=\text{GP}_{\text{mean}}\left( {\widetilde{\mathbf{E}}_n} \right),
\end{equation}
\begin{equation}\label{}
{\mathbf{E}}^\text{std}_{n}=\text{GP}_{\text{std}}\left( {\widetilde{\mathbf{E}}_n} \right).\\
\end{equation}
Note that although the operations, including extracting feature maps by CNN, incorporating saliency map via attention mechanism,
and obtaining statistics of mean and standard deviation,
are the same in both the content-dependency module and the edge masking module, their final outputs correspond to different types of features.
The reason is that the input of CNN in the content-dependency module is the original frame $\mathbf{I}_{n}$, while that in the edge masking module is the processed edge maps $\mathbf{D}_{n}$.
Therefore, the former module is responsible for capturing the content-related features, while the latter module is responsible for capturing the edge-related features.
These two types of features make up the final spatial features in the proposed HVS-5M.


\subsection{Motion Perception Module}

Biological studies on HVS have shown that human M-cells are sensitive to motion changes \cite{M-cell1,M-cell2,M-cell3}.
However, most of the existing VQA models based on extracting frame-level features merely consider the spatial information, and independently process different frames within a video.
To better simulate HVS, the proposed HVS-5M utilizes the motion perception module to process video sequences as a whole, and capture the interframe changes in the temporal domain.

Specifically, SlowFast \cite{SlowFast} pre-trained on Kinetic-400 \cite{Kinetic-400} is adopted to extract the motion feature maps $\mathbf{M}$ from video sequences as
\begin{equation} \label{Motion extraction}
\mathbf{M}=\text{SlowFast}\left( \left\{ \mathbf{I}_n \right\} _{n=1}^{N} \right) \ .
\end{equation}
SlowFast has been widely applied in video recognition.
To simulate human P-cells and M-cells, it operates in slow path at low frame rates to capture spatial semantics and operates in fast path at high frame rates to capture motion at fine temporal resolution.
By this means, temporal information can be obtained without computing the optical flow, and is a supplement to spatial information. Kinetic-400 is a video dataset for action recognition.
Pre-training SlowFast on Kinetic-400 is beneficial to extract significant motion features for video sequences.
Since the spatial features have already been considered in the content-dependency module and the edge masking module,
we only take the feature maps in the `res5c' layer in the fast path as motion feature maps $\mathbf{M}$.
Afterwards, we also perform ($\text{GP}_\text{mean}$) and ($\text{GP}_\text{std}$) on $\mathbf{M}$ to further obtain statistic features.
\begin{equation}\label{}
{\mathbf{M}}^\text{mean}=\text{GP}_{\text{mean}}\left( \mathbf{M} \right),
\end{equation}
\begin{equation}\label{}
{\mathbf{M}}^\text{std}=\text{GP}_{\text{std}}\left( \mathbf{M} \right).
\end{equation}



\begin{table*}[]
\caption{Brief introduction of the six video datasets.}\label{Dataset Discription}
\begin{threeparttable}
\begin{tabular}{ccccccc}
\midrule
\textbf{Dataset}          & KoNViD-1k        & CVD2014            & LIVE-VQC              & LIVE-Qualcomm          & YouTube-UGC        & LSVQ                  \\ \midrule
Number of Videos           & 1200             & 234                & 585                   & 208                    & 1142\tnote{1}                & 39072                 \\
Format                     & RGB              & RGB                & RGB                   & YUV                    & YUV                & RGB                   \\
Framerate                  & 24,25 or 30      & 11 to 31           & 19 to 30 (one 120)    & 30                     & 15 to 60           & -                     \\
Video Duration             & 8                & 10 to 25           & \textgreater{}10      & 15                     & 20                 & 5 to 12               \\
Distortion type            & In-the-wild      & In-capture         & In-the-wild           & In-capture             & In-the-wild        & In-the-wild           \\
Subjective study framework & Crowdsourced     & In-lab             & Crowdsourced          & In-lab                 & Crowdsourced       & Crowdsourced          \\
Annotators                 & 642              & 210                & 4776                  & 39                     & -                  & 6284                  \\
Video Resolutions          & 540p             & 480p, 720p         & 240p-1080p            & 1080p                  & 360p-4k            & 99p-4k                \\
Time duration			& 8s              & 10-25s              & 10s                 & 15s                     & 20s                  & 5-12s
\\
Max length 			& 240              & 830                & 1202                  & 526                     & 2819                  & 4096
\\
Annotation Range           & {[}1.22, 4.64{]} & {[}-6.50, 93.38{]} & {[}6.2237, 94.2865{]} & {[}16.5621, 73.6428{]} & {[}1.242, 4.698{]} & {[}2.4483, 91.4194{]} \\ \hline

\end{tabular}
\begin{tablenotes}
        \footnotesize
        \item[1] After removing 57 grayscale videos, the latest version of YouTube-UGC has 1142 videos.
      \end{tablenotes}
\end{threeparttable}
\end{table*}

\subsection{Temporal Hysteresis Module}

In HVS, human beings are more likely to remember those frames with poor quality and assign such video with a low quality score, even if the amount of frames with poor quality is quite small \cite{Temporal-Memory}.
Therefore, in the proposed HVS-5M, the temporal hysteresis module is applied to simulate such memory mechanism.

The proposed HVS-5M formulates a given video in both spatial domain and temporal domain.
Correspondingly, the temporal hysteresis module
firstly fuses various types of features.
From the perspective of spatial domain, the content features (${\mathbf{C}^\text{mean}_{n}}$, ${\mathbf{C}^\text{std}_{n}}$) and the edge features (${\mathbf{E}^\text{mean}_{n}}$, ${\mathbf{E}^\text{std}_{n}}$) are concatenated as spatial features $\mathbf{S}_{n}$ as
\begin{equation}\label{}
\mathbf{S}_{n}=\text{Concat}\left( \mathbf{C}^\text{mean}_{n},\mathbf{C}^\text{std}_{n},\mathbf{E}^\text{mean}_{n},\mathbf{E}^\text{mean}_{n} \right).
\end{equation}
From the perspective of temporal domain, the motion features (${\mathbf{M}}^\text{mean}$, ${\mathbf{M}}^\text{std}$) are concatenated as temporal features  $\mathbf{T}$ as
\begin{equation}\label{}
\mathbf{T}=\text{Concat}\left( {\mathbf{M}}^\text{mean}, {\mathbf{M}}^\text{std} \right) .
\end{equation}
Then, $\mathbf{S}_{n}$ and $\mathbf{T}$ are integrated into fusion features $\mathbf{F}_n$ as
\begin{equation}\label{}
\mathbf{F}_n=\text{Concat}\left( \text{Sample}\left( \left\{ \mathbf{S}_n \right\}_{n=1}^{N}\right), \mathbf{T}\right).
\end{equation}
wherein $\left\{ \mathbf{S}_n \right\}_{n=1}^{N}$ are firstly sampled in every two frames so that
the first dimension of {\color{blue}}$\text{Sample}\left( \left\{ \mathbf{S}_n \right\}_{n=1}^{N} \right)$ and $\mathbf{T}$ are the same, and then $\text{Sample}\left( \left\{ \mathbf{S}_n \right\}_{n=1}^{N} \right)$ and $\mathbf{T}$ can be concatenated along the second dimension.


Secondly, the temporal hysteresis module predicts the frame-level quality score.
To remove the redundant information, $\mathbf{F}_n$ is processed by fully-connected network as
\begin{equation}\label{}
\hat{\mathbf{F}}_n=\text{FC}\left( \mathbf{F}_{n} \right).
\end{equation}
Then, the module utilizes TempHyst \cite{Temporal-Memory} to obtain the frame-level quality score $q_n$ based on current $\hat{\mathbf{F}}_n$ and previous $\hat{\mathbf{F}}_{n-1}$ via GRU network \cite{GRU} as
\begin{equation}\label{}
q_n=\text{FC}\left( \text{GRU}\left( \hat{\mathbf{F}}_n \right) ,\text{GRU}\left( \hat{\mathbf{F}}_{n-1} \right) \right).
\end{equation}

Thirdly, TempHyst aims to highlight those frames with poor quality within video sequences.
Considering $n$ as the current time step,
the quality score of the worst frame within max(1, $n-\tau$) to $(n-1)$
is denoted as $x_n$, and is calculated as
\begin{equation}\label{}
x_{n}=\left\{
             \begin{array}{lr}
            q_n, & n = 1, \\
             \mathop{\text{min}\ q_k}\limits_{k\in V_{pre}},  &  n>1,\\
             \end{array}
\right.
\end{equation}
where $V_{pre} $=$ \left\{ \text{max} (1,n-\tau), ..., n-2, n-1 \right\}$, and $\tau$ is the hyperparameter.
On the other hand, considering $n$ as the current time step,
the weighted average of quality scores within $n$ to min($n+\tau,N$) is denoted as $y_n$, and is calculated as
\begin{equation}\label{}
y_n = \sum_{k\in V_{next}}\ q_kw^k_n,
\end{equation}
where $w^k_n$ represents the contribution of $q_k$ towards $y_n$ and is calculated as
\begin{equation}\label{}
w^k_n = \frac{e^{-q_k}}{\sum_{j\in V_{next}}e^{-q_j}}, k\in V_{next},
\end{equation}
\noindent
where $V_{next} $=$ \left\{ n, n+1, ..., \text{min}(n+\tau,N)\right\} $, and  $\tau$ is the hyperparameter.

Fourthly, TempHyst calculates the frame quality score $q_n^{\prime}$ as the linear combination of $x_n$ and $y_n$ as
\begin{equation}\label{}
q_n^{\prime} = \gamma x_n + ( 1 - \gamma)y_n,
\end{equation}
where $\gamma$ is the hyperparameter that determines the contribution of $x_n$ and $y_n$.
Finally, the overall video quality score $Q$ can be calculated as the
the average of the frame quality scores within the whole video sequences as
\begin{equation}\label{}
Q = \frac{1}{N} \sum^{N}_{n=1}q_n^{\prime}.
\end{equation}

\section{Experimental Results and Analysis}
\label{sec:exp}

The organizations of Section \ref{sec:exp} are as follows. The experimental settings are firstly described (given in Section \ref{experimental settings}). Then, in order to evaluate the performance of the proposed HVS-5M, experiments are conducted under different settings, including on individual datasets (given in Section \ref{individual datasets}), on categorical subsets (given in Section \ref{categorical subsets}), on cross datasets (given in Section \ref{cross datasets}), and on mixed datasets (given in Section \ref{mixed datasets}). Finally, ablation studies are conducted to verify the effectiveness of different modules (given in Section \ref{ablation studies}).

\begin{table*}[]
\caption{Performance evaluations on individual datasets. Here, the top performer is highlighted in boldface, and N/A indicates ``Not applicable''.}\label{Individual Dataset}
\centering
\fontsize{9.2}{9.2}\selectfont
\subtable[Performance evaluations on CVD2014 \cite{CVD2014} and LIVE-Qualcomm \cite{LIVE-Qualcomm}.]{
\begin{tabular}{ccccc}
\toprule[1.1pt]
\multirow{2}{*}{Type}   & \multirow{2}{*}{Method} & CVD2014   &  & LIVE-Qualocmm  \\ \cmidrule(lr){3-3} \cmidrule(lr){5-5}
                        &                         & SRCC   \quad  PLCC   &  & SRCC  \qquad   PLCC              \\ \midrule
\multirow{4}{*}{IQA} & NIQE \cite{NIQE}                 & 0.4755 \quad 0.6070  &  & 0.4608 \qquad 0.5336            \\
                        & BRISQUE \cite{BRISQUE}                  & 0.7900 \quad 0.8049  &  & 0.5527 \qquad 0.5986     \\
                        & FRIQUEE \cite{FRIQUEE}                  & 0.8212 \quad 0.8415  &  & 0.7158 \qquad 0.7481      \\
                        & CORNIA \cite{CORNIA}                  & 0.6277 \quad 0.6631  &  & 0.4551 \qquad 0.5203           \\
						& CONTRIQUE \cite{CONTRIQUE}                  & 0.826 \quad 0.836  &  & 0.765 \qquad 0.777     \\
\midrule
\multirow{7}{*}{VQA} & VIIDEO \cite{VIIDEO}                  & 0.0503 \quad 0.2479 &  & 0.0808 \qquad 0.2301        \\
                        & V-BLIINDS \cite{VBLIINDS}               & 0.7950 \quad 0.8067  &  & 0.5702 \qquad 0.6269      \\
                        & TLVQM  \cite{TLVQM}                  & 0.7799 \quad 0.7904    &  & 0.7849 \qquad 0.8152          \\

                        & VSFA      \cite{VSFA}               & 0.8501 \quad 0.8690  &  & 0.7080 \qquad 0.7741       \\
					
                        & BVQA-2022       \cite{Li2022}             & 0.8626 \quad 0.8826  &  & \textbf{0.8334} \qquad \textbf{0.8371}
      \\
						& CONVIQT       \cite{CONVIQT}             & 0.858 \quad 0.837  &  & 0.797 \qquad 0.802      \\
                        & HVS-5M         &     \textbf{0.8729}  \quad        \textbf{0.8853} &  &   0.8090  \qquad        0.8210             \\ \bottomrule[1.1pt]
\end{tabular}}

~\\
~\\

\subtable[Performance evaluations on KoNViD-1k \cite{KoNViD-1k}, LIVE-VQC \cite{LIVE-VQC}, and YouTube-UGC \cite{YouTube-UGC}.]{
\centering
\fontsize{9.2}{9.2}\selectfont
\setlength{\tabcolsep}{1.4mm}{
\begin{tabular}{cccccccc}
\toprule[1.1pt]
\multirow{2}{*}{Type}    & \multirow{2}{*}{Method} & KoNViD-1k &  & LIVE-VQC   &  & YouTube-UGC \\ \cmidrule(lr){3-3} \cmidrule(lr){5-5} \cmidrule(lr){7-7}
                         &                         & SRCC \quad   PLCC    &  & SRCC  \quad  PLCC    &  & SRCC \qquad   PLCC             \\ \midrule
\multirow{12}{*}{IQA} & NIQE \cite{NIQE}                   & 0.5392 \quad 0.5513   &  & 0.5930 \quad 0.6312   &  & 0.2499 \qquad 0.2982             \\
                         & IL-NIQE  \cite{IL-NIQE}               & 0.5199 \quad 0.5371   &  & 0.5019 \quad 0.5433   &  & 0.3198 \qquad 0.3585             \\
                         & BRISQUE  \cite{BRISQUE}               & 0.6493 \quad 0.6513   &  & 0.5936 \quad 0.6242   &  & 0.3932 \qquad 0.4073             \\
                         & GM-LOG    \cite{GM-LOG}                   & 0.6422 \quad 0.6452   &  & 0.5876 \quad 0.6218   &  & 0.3450 \qquad 0.3769             \\
                         & HIGRADE \cite{HIGRADE}                & 0.7062 \quad 0.7104   &  & 0.5959 \quad 0.6188   &  & 0.7252 \qquad 0.7103              \\
                         & FRIQUEE  \cite{FRIQUEE}               & 0.7352 \quad 0.7354   &  & 0.6502 \quad 0.6914   &  & 0.7538 \qquad 0.7505              \\
                         & CORNIA   \cite{CORNIA}               & 0.7351 \quad 0.7356   &  & 0.6808 \quad 0.7239   &  & 0.5671 \qquad 0.5851              \\
                         & HOSA     \cite{HOSA}               & 0.7606 \quad 0.7580   &  & 0.6784 \quad 0.7242   &  & 0.5961 \qquad 0.6037               \\
                         & VGG-19    \cite{VGG}              & 0.7209 \quad 0.7385   &  & 0.6762 \quad 0.7281   &  & 0.6037 \qquad 0.6074              \\
                         & ResNet-50  \cite{ResNet}             & 0.7651 \quad 0.7781   &  & 0.6814 \quad 0.7381   &  & 0.6542 \qquad 0.6485              \\
                         & KonCept512 \cite{KonCept512}             & 0.7349 \quad 0.7489   &  & 0.6645 \quad 0.7278   &  & 0.5872 \qquad 0.5940              \\
                         & PaQ-2-PiQ   \cite{PaQ-2-PiQ}            & 0.6130 \quad 0.6014   &  & 0.6436 \quad 0.6683   &  & 0.2658 \qquad 0.2935              \\
						 & CONTRIQUE   \cite{CONTRIQUE}               & 0.844 \quad 0.842   &  & 0.815 \quad 0.822   &  & 0.825 \qquad 0.813              \\
 \midrule
\multirow{9}{*}{VQA}  & VIIDEO   \cite{VIIDEO}               & 0.2874 \quad 0.3083   &  & 0.0461 \quad 0.2100   &  & 0.0567 \qquad 0.1497               \\
                         & V-BLIINDS  \cite{VBLIINDS}             & 0.7063 \quad 0.7011   &  & 0.6811 \quad 0.6997   &  & 0.5348 \qquad 0.5409            \\
                         & TLVQM    \cite{TLVQM}               & 0.7588 \quad 0.7598   &  & 0.7878 \quad 0.7942   &  & 0.6568 \qquad 0.6470               \\
                         & VIDEVAL  \cite{VIDEVAL}               & 0.7704 \quad 0.7709 &  & 0.7438 \quad 0.7476 &  & 0.7763 \qquad 0.7715           \\
                         & VSFA    \cite{VSFA}                & 0.7943 \quad 0.7985   &  & 0.7176 \quad 0.7707   &  & 0.7873 \qquad 0.7888              \\
                         & RAPIQUE  \cite{RAPIQUE}               & 0.7884 \quad 0.8051 &  & 0.7413 \quad 0.7618 &  & 0.7473 \qquad 0.7569           \\
                         & PVQ    \cite{LSVQ}                 & 0.791 \quad 0.786   &  & 0.827 \quad 0.837   &  & N/A    \qquad  N/A                          \\
					
                         & BVQA-2022    \cite{Li2022}               & 0.8354 \quad 0.8339   &  & 0.8414 \quad 0.8394   &  & 0.8252 \qquad 0.8178              \\
						 & CONVIQT    \cite{CONVIQT}               & 0.851 \quad 0.849   &  & 0.808 \quad 0.817   &  & 0.832 \qquad 0.822              \\
                         & HVS-5M        &       \textbf{0.8530}  \quad        \textbf{0.8562}          &  &     \textbf{0.8441}  \quad        \textbf{0.8422}            &  &     \textbf{0.8520}  \qquad        \textbf{0.8451}                                  \\ \bottomrule[1.1pt]
\end{tabular}}}

~\\
~\\

\subtable[Performance evaluations on LSVQ \cite{LSVQ}.]{
\centering
\fontsize{9.2}{9.2}\selectfont
\begin{tabular}{cccccc}
\toprule[1.1pt]
\multirow{2}{*}{Type} & \multirow{2}{*}{Method} & Test    &  & Test-1080p \\ \cmidrule(lr){3-3} \cmidrule(lr){5-5}
                      &                         & SRCC \quad   PLCC  &  & SRCC  \quad  PLCC               \\ \midrule
\multirow{2}{*}{IQA}                  & BRISQUE  \cite{BRISQUE}                & 0.579 \quad 0.576   &  & 0.497 \quad 0.531                \\
					  & CONTRIQUE  \cite{CONTRIQUE}                & 0.828 \quad 0.826   &  & 0.662 \quad 0.697                \\
 \midrule
\multirow{7}{*}{VQA}  & TLVQM   \cite{TLVQM}                 & 0.772 \quad 0.774   &  & 0.589 \quad 0.616                \\
                      & VIDEVAL  \cite{VIDEVAL}                & 0.794 \quad 0.783   &  & 0.545 \quad 0.554               \\
                      & VSFA    \cite{VSFA}                 & 0.801 \quad 0.796   &  & 0.675 \quad 0.704               \\
                      & PVQ (w/o)  \cite{LSVQ}               & 0.814 \quad 0.816 &  & 0.686 \quad 0.708               \\
                      & PVQ (w)  \cite{LSVQ}                 & 0.827 \quad 0.828 &  & 0.711 \quad 0.739               \\
                      & BVQA-2022  \cite{Li2022}                  & 0.8518 \quad 0.8535 &  & 0.7718 \quad 0.7878             \\
 					  & CONVIQT  \cite{CONVIQT}                  & 0.821 \quad 0.820 &  & 0.668 \quad 0.702             \\
                      & HVS-5M         &      \textbf{0.8785}  \quad        \textbf{0.8723}         &  &    \textbf{0.7977}  \quad        \textbf{0.8146}                            \\ \bottomrule[1.1pt]
\end{tabular}}
\end{table*}

\subsection{Experimental Setup} \label{experimental settings}

\subsubsection{Compared methods}
Our HVS-5M is compared with 25 leading methods, including VQA methods such as VIIDEO \cite{VIIDEO}, V-BLIINDS \cite{VBLIINDS}, TLVQM \cite{TLVQM}, VIDEVAL \cite{VIDEVAL}, CNN-TLVQM \cite{CNN-TLVQM}, VSFA \cite{VSFA}, GSTVQA \cite{Chen2021}, MDTVSFA \cite{MDTVSFA}, RAPIQUE \cite{RAPIQUE}, PVQ \cite{LSVQ}, BVQA-2022 \cite{Li2022}, CONVIQT \cite{CONVIQT}, and image quality assessment (IQA) methods including NIQE \cite{NIQE}, IL-NIQE \cite{IL-NIQE}, BRISQUE \cite{BRISQUE}, GM-LOG \cite{GM-LOG}, HIGRADE \cite{HIGRADE}, FRIQUEE \cite{FRIQUEE}, CORNIA \cite{CORNIA}, HOSA \cite{HOSA}, CONTRIQUE \cite{CONTRIQUE}, VGG-19 \cite{VGG}, ResNet-50 \cite{ResNet}, KonCept512 \cite{KonCept512}, PaQ-2-PiQ \cite{PaQ-2-PiQ}.
We collect the experimental results of the compared methods from \cite{Chen2021,MDTVSFA,Li2022,CONVIQT}.
\begin{table}[]
\caption{Performance evaluations on categorical subsets according to resolution.}\label{resolution}
\centering
\fontsize{9.2}{9.2}\selectfont
\setlength{\belowcaptionskip}{-0.5cm}
\setlength{\tabcolsep}{0.85mm}{
\begin{tabular}{cccc}
\toprule[1.1pt]
\multirow{2}{*}{Method} & 1080p         &   720p          &   $\leq$ 480p \\ \cmidrule(lr){2-2} \cmidrule(lr){3-3} \cmidrule(lr){4-4}
                        & SRCC \   PLCC    & SRCC \   PLCC    & SRCC  \  PLCC     \\ \midrule
BRISQUE \cite{BRISQUE}                 & 0.4447 \ 0.4645   & 0.5972 \ 0.5950   & 0.5092 \ 0.5088    \\
GM-LOG  \cite{GM-LOG}                   & 0.4541 \ 0.4958   & 0.5129 \ 0.4933   & 0.4989 \  0.5180    \\
HIGRADE  \cite{HIGRADE}             & 0.4574 \ 0.5316   & 0.5447 \   0.5478   & 0.6147 \ 0.6282    \\
FRIQUEE \cite{FRIQUEE}            & 0.5513 \ 0.6039   & 0.6130 \ 0.6118 &   0.6735 \ 0.6966    \\
CORNIA  \cite{CORNIA}               & 0.6140 \ 0.7063   & 0.6176 \ 0.6190   & 0.6667 \ 0.7080    \\
HOSA  \cite{HOSA}                & 0.5747 \ 0.6423   & 0.6578 \ 0.6489   & 0.7158 \ 0.7302    \\
VGG-19    \cite{VGG}              & 0.6240 \ 0.6546   & 0.6407 \ 0.6370   & 0.7076 \ 0.7138    \\
ResNet-50  \cite{ResNet}             & 0.6373 \ 0.6713   & 0.6548 \ 0.6781   & 0.7591 \ 0.7745    \\
V-BLIINDS  \cite{VBLIINDS}             & 0.4048 \ 0.5063   & 0.5642 \ 0.5758   & 0.5912 \ 0.6096    \\
TLVQM    \cite{TLVQM}              & 0.5530 \ 0.6329  & 0.6464 \ 0.6467   & 0.6113 \ 0.6332    \\
VIDEVAL \cite{VIDEVAL}               & 0.5536 \ 0.6016   & 0.6312 \ 0.6381   & 0.6106 \ 0.6667    \\
RAPIQUE \cite{RAPIQUE}               & 0.5262 \ 0.6230   & 0.5911 \ 0.6362   & 0.6820 \ 0.6927    \\
VSFA     \cite{VSFA}           & 0.6512 \ 0.6697   & 0.6677 \ 0.6650   & 0.7084 \ 0.7277    \\
BVQA-2022  \cite{Li2022}                 & 0.6759 \ 0.6902   & 0.7431 \ 0.7333   & 0.7819 \ 0.7972    \\
HVS-5M        &       \textbf{0.7485} \  \textbf{0.7687}        &                \textbf{0.7566} \  \textbf{0.7518}  &  \textbf{0.7834} \  \textbf{0.8023} \\ \bottomrule[1.1pt]
\end{tabular}}
\end{table}


\begin{table}[]
\caption{Performance evaluations on categorical subsets according to content.}\label{content}
\centering
\fontsize{9.2}{9.2}\selectfont
\setlength{\tabcolsep}{0.85mm}{
\begin{tabular}{cccc}
\toprule[1.1pt]
\multirow{2}{*}{Method} & Screen Content &   Animation     &   Gaming        \\ \cmidrule(lr){2-2} \cmidrule(lr){3-3} \cmidrule(lr){4-4}
                        & SRCC \   PLCC   &   SRCC \   PLCC  &   SRCC \   PLCC  \\ \midrule
BRISQUE   \cite{BRISQUE}               & 0.1770 \ 0.3937  &   0.0945 \ 0.4862 &   0.3200 \ 0.3605 \\
GM-LOG   \cite{GM-LOG}                   & 0.2167 \ 0.2387  &   0.2813 \ 0.5212 &   0.2505 \ 0.3262 \\
HIGRADE   \cite{HIGRADE}               & 0.4603 \ 0.5555  &   0.2703 \ 0.4348 &   0.5692 \ 0.6233 \\
FRIQUEE   \cite{FRIQUEE}               & 0.4779 \ 0.5612  &   0.1846 \ 0.4929 &   0.6527 \ 0.6944 \\
CORNIA   \cite{CORNIA}                & 0.3304 \ 0.4187  &   0.2461 \ 0.4674 &   0.5241 \ 0.6060 \\
HOSA    \cite{HOSA}                 & 0.3396 \ 0.5127  &   0.1329 \ 0.4591 &   0.5611 \ 0.6057 \\
VGG-19   \cite{VGG}                & 0.4025 \ 0.4689  &   0.1340 \ 0.4182 &   0.5436 \ 0.5841 \\
ResNet-50 \cite{ResNet}               & 0.4213 \ 0.5220  &   0.2549 \ 0.4270 &   0.5639 \ 0.5945 \\
V-BLIINDS \cite{VBLIINDS}               & 0.1730 \ 0.3036  &   -0.1560 \ 0.4471 &   0.4138 \ 0.5447 \\
TLVQM     \cite{TLVQM}               & 0.3266 \ 0.4690  &   0.1274 \ 0.3926 &   0.6014 \ 0.6213 \\
VIDEVAL  \cite{VIDEVAL}                & 0.5307 \ 0.6469  &   0.2109 \ 0.4480 &   0.6971 \ 0.7105 \\
RAPIQUE   \cite{RAPIQUE}               & 0.4392 \ 0.4514  &   0.3120 \ 0.4882 &   0.5471 \ 0.6643 \\
VSFA      \cite{VSFA}            & 0.4700 \ 0.5493  &   -0.0021 \ 0.4355 &   0.6774 \ 0.7501 \\
BVQA-2022   \cite{Li2022}                 & 0.5726 \ 0.6845  &   0.5604 \ \ 0.6635 &   0.7073 \ 0.7655 \\
HVS-5M               &   \textbf{0.7495} \  \textbf{0.7155}             &                \textbf{0.6423} \  \textbf{0.7235} &    \textbf{0.8632} \  \textbf{0.8505}              \\ \bottomrule[1.1pt]
\end{tabular}}
\end{table}


\begin{table}[]
\caption{Performance evaluations on categorical subsets according to quality.}\label{quality}
\centering
\fontsize{9.1}{9.2}\selectfont

\begin{tabular}{cccc}
\toprule[1.1pt]
\multirow{2}{*}{Method} & Low Quality   &  & High Quality  \\ \cmidrule(lr){2-2} \cmidrule(lr){4-4}
                        & SRCC \quad   PLCC  &  & SRCC \quad   PLCC  \\ \midrule
BRISQUE   \cite{BRISQUE}              & 0.4347 \quad 0.4617 &  & 0.2986 \quad 0.3110 \\
GM-LOG    \cite{GM-LOG}                  & 0.3992 \quad 0.4840 &  & 0.2384 \quad 0.2328 \\
HIGRADE    \cite{HIGRADE}             & 0.5463 \quad 0.5707 &  & 0.4659 \quad 0.4792 \\
FRIQUEE   \cite{FRIQUEE}              & 0.5319 \quad 0.5926 &  & 0.5004 \quad 0.5360 \\
CORNIA    \cite{CORNIA}              & 0.4867 \quad 0.5454 &  & 0.3619 \quad 0.3712 \\
HOSA      \cite{HOSA}              & 0.5411 \quad 0.5874 &  & 0.4296 \quad 0.4402 \\
VGG-19     \cite{VGG}             & 0.5528 \quad 0.5969 &  & 0.4332 \quad 0.4375 \\
ResNet-50  \cite{ResNet}             & 0.5944 \quad 0.6476 &  & 0.4621 \quad 0.4701 \\
V-BLIINDS  \cite{VBLIINDS}             & 0.5033 \quad 0.5095 &  & 0.3503 \quad 0.3545 \\
TLVQM      \cite{TLVQM}             & 0.5026 \quad 0.5497 &  & 0.5064 \quad 0.5130 \\
VIDEVAL   \cite{VIDEVAL}              & 0.5816 \quad 0.6194 &  & 0.5558 \quad 0.5830 \\
RAPIQUE    \cite{RAPIQUE}             & 0.6349 \quad 0.6828 &  & 0.5250 \quad 0.5391 \\
VSFA       \cite{VSFA}          & 0.6169 \quad 0.6741 &  & 0.5286 \quad 0.5313 \\
BVQA-2022    \cite{Li2022}              & 0.6863 \quad 0.7118 &  & \textbf{0.6524} \quad \textbf{0.6694} \\
HVS-5M                &    \textbf{0.6900} \quad  \textbf{0.7322}            &  &    0.6410 \quad 0.6425           \\ \bottomrule[1.1pt]
\end{tabular}
\end{table}


\subsubsection{Datasets}
The performance of different methods is evaluated on six mainstream video datasets, including KoNViD-1k \cite{KoNViD-1k}, CVD2014 \cite{CVD2014},  LIVE-VQC \cite{LIVE-VQC}, LIVE-Qualcomm \cite{LIVE-Qualcomm}, YouTube-UGC \cite{YouTube-UGC}, and LSVQ \cite{LSVQ}. Their brief information is given in Table \ref{Dataset Discription}.
As for LSVQ, we select two subsets for training and testing.
As for the other five datasets, we randomly split the dataset into training set, validation set, and testing set with a proportion of 60$\%$, 20$\%$, and 20$\%$.

In order to reduce the numerical errors when computing the color model FRIQUEE \cite{FRIQUEE},
we remove 57 grayscale videos in YouTube-UGC \cite{YouTube-UGC}, as it did in \cite{VIDEVAL}.

\subsubsection{Performance metrics}
The performance of different methods is evaluated by Spearman Rank order Correlation Coefficient (SRCC) and Pearson's Linear Correlation Coefficient (PLCC).
Specifically, SRCC is utilized to measure the monotonic relationship between the ground truth and the predicted quality scores. And PLCC is utilized to measure the accuracy of predictions.




\subsubsection{Implementation details}
The loss function of our HVS-5M is defined based on
SRCC and PLCC as
\begin{equation}\label{}
\mathscr{L}=\mathscr{L}_{\text{SRCC}}+ \mathscr{L}_{\text{PLCC}}.
\end{equation}
Specifically,
\begin{equation}\label{}
\mathscr{L}_{\text{SRCC}}=1-\text{SRCC},
\end{equation}
\begin{equation}\label{}
\text{SRCC}=\frac{\sum_{i=1}^N{\left( y_{r}^{_i}-\bar{y}_r \right)}\left( \hat{y}_{r}^{i}-\hat{\bar{y}}_r \right)}{\sqrt{\sum_{i=1}^N{\left( y_{r}^{i}-\bar{y}_r \right)}^2}\sqrt{\sum_{i=1}^N{\left( \hat{y}_{r}^{i}-\hat{\bar{y}}_r \right)}^2}},
\end{equation}
where SRCC is a differentiable approximation of its original metric,
$\bar{y}_r$ and $\hat{\bar{y}}_r$ denote the ranks of the ground truth and the predicted quality scores, respectively.
\begin{equation}\label{}
\mathscr{L}_{\text{PLCC}}=\left( 1-\mathrm{PLCC} \right) /2,
\end{equation}
\begin{equation}\label{}
	\mathrm{PLCC}=\frac{\begin{matrix} \sum_{i=1}^N (y_{i}-\bar{y})(\hat{y_{i}}-\hat{\bar{y}}) \end{matrix}}{\sqrt{\begin{matrix} \sum_{i=1}^N (y_{i}-\bar{y})^{2}\end{matrix}} \sqrt{\begin{matrix} \sum_{i=1}^N (\hat{y_{i}}-\hat{\bar{y}})^{2}\end{matrix}}},
\end{equation}
\noindent where $\bar{y}$ and $\hat{\bar{y}}$ denote the mean value of the ground truth and the predicted quality scores, respectively.

Our proposed HVS-5M is built on Pytorch 3.8 and trained on two GeForce RTX A10 GPU cards. SAMNet \cite{SAMNet}, ConvNeXt-XL \cite{ConvNeXt}, and SlowFast \cite{SlowFast} are adopted in HVS-5M.
The dimension of the fusion features is reduced to 256 via fully-connected layer before being fed into the GRU network \cite{GRU}, and the hidden size of the GRU unit is set to 72.
$\tau$ and $\gamma$ in the temporal hysteresis module are set to 12 and 0.5, respectively.
Adam optimizer \cite{Adam} is used for optimization, wherein the learning rate is initialized as $10^{-5}$ and decayed by 0.2 for every two epochs.
The results are averaged over 10 runs under the same setup.

\renewcommand\arraystretch{1.15}
\begin{table*}[]
\caption{Performance evaluations on cross datasets.}\label{Cross-Dataset1}
\centering
\fontsize{9.2}{9.2}\selectfont
\setlength{\tabcolsep}{0.9mm}{
\begin{tabular}{c|ccccc|ccccc}
\Xhline{1.1pt}
Training                 & \multicolumn{5}{c|}{KoNViD-1k}                        & \multicolumn{5}{c}{LIVE-Qualcomm}                    \\ \hline
\multirow{2}{*}{Testing} & CVD2014        &  & LIVE-Qualcomm &  & LIVE-VQC      & KoNViD-1k      &  & CVD2014       &  & LIVE-VQC      \\   \cline{2-2} \cline{4-4} \cline{6-7} \cline{9-9} \cline{11-11}
                         & SRCC \quad   PLCC   &  & SRCC \quad   PLCC  &  & SRCC  \quad  PLCC  & SRCC \quad   PLCC   &  & SRCC \quad   PLCC  &  & SRCC \quad   PLCC  \\ \hline
NIQE    \cite{NIQE}                 & 0.3856 \quad 0.4410  &  & 0.1807 \quad 0.1672 &  & 0.4573 \quad 0.4025 & 0.4564 \quad 0.3619  &  & 0.3856 \quad 0.4410 &  & 0.4573 \quad 0.4025 \\
BRISQUE   \cite{BRISQUE}               & 0.4626 \quad 0.5060  &  & 0.3061 \quad 0.3303 &  & 0.5805 \quad 0.5788 & 0.4370 \quad 0.4274  &  & 0.4626 \quad 0.5060 &  & 0.5805 \quad 0.5788 \\
VSFA     \cite{VSFA}                & 0.6278 \quad 0.6216  &  & 0.5574 \quad 0.5769 &  & 0.6792 \quad 0.7198 & 0.6643 \quad 0.6116  &  & 0.5348 \quad 0.5606 &  & 0.6425 \quad 0.6819 \\
TLVQM     \cite{TLVQM}               & 0.3569 \quad 0.3838  &  & 0.4730 \quad 0.5127 &  & 0.5953 \quad 0.6248 & 0.0347 \quad 0.0467  &  & 0.4893 \quad 0.4721 &  & 0.4091 \quad 0.3559 \\
VIDEVAL   \cite{VIDEVAL}               & 0.6494 \quad 0.6638  &  & 0.4048 \quad 0.4351 &  & 0.5318 \quad 0.5329 & 0.1812 \quad -0.3441 &  & 0.6059 \quad 0.6244 &  & 0.4314 \quad 0.4122 \\
CNN-TLVQM  \cite{CNN-TLVQM}              & 0.6828 \quad 0.7226  &  & 0.6050 \quad 0.6223 &  & 0.7132 \quad 0.7522 & 0.0854 \quad 0.0216  &  & 0.2367 \quad 0.2388 &  & 0.0693 \quad 0.1040 \\
GSTVQA    \cite{Chen2021}             & 0.7972 \quad 0.7984  &  & 0.6200 \quad 0.6666 &  & 0.6797 \quad 0.7327 & 0.6694 \quad 0.6258  &  & \textbf{0.7046} \quad 0.6665 &  & 0.6201 \quad 0.6100   \\
HVS-5M                 &   \textbf{0.8042} \quad  \textbf{0.8097}             &  &   \textbf{0.6927} \quad  \textbf{0.7179}             &  &              \textbf{0.7728} \quad  \textbf{0.7995}  &      \textbf{0.7100} \quad  \textbf{0.6810}           &  &   0.7017 \quad  \textbf{0.6987}             &  &   \textbf{0.7626} \quad  \textbf{0.7577}             \\ \hline  \hline
Training                 & \multicolumn{5}{c|}{LIVE-VQC}                         & \multicolumn{5}{c}{CVD2014}                          \\ \hline
\multirow{2}{*}{Testing} & KoNViD-1k      &  & CVD2014       &  & LIVE-Qualcomm & KoNViD-1k      &  & LIVE-Qualcomm &  & LIVE-VQC      \\ \cline{2-2} \cline{4-4} \cline{6-7} \cline{9-9} \cline{11-11}
                         & SRCC \quad   PLCC   &  & SRCC \quad   PLCC  &  & SRCC \quad   PLCC  & SRCC  \quad  PLCC   &  & SRCC   \quad  PLCC  &  & SRCC  \quad  PLCC  \\ \hline
NIQE     \cite{NIQE}                & 0.4564 \quad 0.3619  &  & 0.3856 \quad 0.4410 &  & 0.1807 \quad 0.1672 & 0.4564 \quad 0.3619  &  & 0.1807 \quad 0.1672 &  & 0.4573 \quad 0.4025 \\
BRISQUE  \cite{BRISQUE}                & 0.4370 \quad 0.4274  &  & 0.4626 \quad 0.5060 &  & 0.3601 \quad 0.3303 & 0.4370 \quad 0.4274  &  & 0.3061 \quad 0.3303 &  & 0.5805 \quad 0.5788 \\
VSFA    \cite{VSFA}                 & 0.6584 \quad 0.6666  &  & 0.5061 \quad 0.5415 &  & 0.5094 \quad 0.5350 & 0.5759 \quad 0.5636  &  & 0.3256 \quad 0.3718 &  & 0.4600 \quad 0.4783 \\
TLVQM    \cite{TLVQM}                & 0.6023 \quad 0.5943  &  & 0.4553 \quad 0.4749 &  & 0.6415 \quad 0.6534 & 0.5437 \quad 0.5052  &  & 0.3334 \quad 0.3838 &  & 0.5397 \quad 0.5527 \\
VIDEVAL   \cite{VIDEVAL}               & 0.5007 \quad -0.4841 &  & 0.5702 \quad 0.5171 &  & 0.3021 \quad 0.3602 & 0.1918 \quad -0.3260 &  & 0.1208 \quad 0.3315 &  & 0.4751 \quad 0.5167 \\
CNN-TLVQM   \cite{CNN-TLVQM}             & 0.6431 \quad 0.6304  &  & 0.6300 \quad 0.6568 &  & 0.6574 \quad 0.6696 & 0.5779 \quad 0.5489  &  & 0.4410 \quad 0.4712 &  & 0.5209 \quad 0.5592 \\
GSTVQA  \cite{Chen2021}               & 0.7085 \quad 0.7074  &  & 0.6894 \quad 0.6645 &  & 0.5952 \quad 0.6245 & 0.6230 \quad 0.5764  &  &  0.4187 \quad 0.4965 &  & 0.5817 \quad 0.5751 \\
HVS-5M                 &  \textbf{0.7487} \quad  \textbf{0.7317}              &  &   \textbf{0.7509} \quad  \textbf{0.7508}            &  &   \textbf{0.7446} \quad  \textbf{0.7239}            &     \textbf{0.6979} \quad  \textbf{0.6788}           &  &   \textbf{0.5791} \quad  \textbf{0.5664          } &  &              \textbf{0.6745} \quad  \textbf{0.6929} \\ \Xhline{1.1pt}
\end{tabular}}
\end{table*}

\begin{table}[]
\caption{Performance evaluations on cross datasets.}\label{Cross-Dataset2}
\centering
\fontsize{9.2}{9.2}\selectfont
\begin{tabular}{cccc}
\toprule[1.1pt]
Training                 & \multicolumn{3}{c}{LSVQ}         \\ \midrule
\multirow{2}{*}{Testing} & KoNViD-1k     &  & LIVE-VQC      \\ \cmidrule(lr){2-2} \cmidrule(lr){4-4}
                         & SRCC \quad   PLCC  &  & SRCC \quad   PLCC  \\ \midrule
BRISQUE  \cite{BRISQUE}                & 0.646 \quad 0.647   &  & 0.524 \quad 0.536   \\
TLVQM    \cite{TLVQM}                & 0.732 \quad 0.724   &  & 0.670 \quad 0.691   \\
VIDEVAL   \cite{VIDEVAL}               & 0.751 \quad 0.741   &  & 0.630 \quad 0.640   \\
VSFA     \cite{VSFA}                & 0.784 \quad 0.794   &  & 0.734 \quad 0.772   \\
PVQ (w/o)  \cite{LSVQ}               & 0.782 \quad 0.781 &  & 0.747 \quad 0.776 \\
PVQ (w)   \cite{LSVQ}                & 0.791 \quad 0.795 &  & 0.770 \quad 0.807\\
BVQA-2022  \cite{Li2022}                  & 0.839 \quad 0.830   &  &   \textbf{0.816} \quad 0.824
            \\
HVS-5M                 &     \textbf{0.857} \quad \textbf{0.855} &  &              0.810 \quad \textbf{0.832} \\ \bottomrule[1.1pt]
\end{tabular}
\end{table}


\begin{table*}[]
\caption{Performance evaluations on mixed datasets. ``KoN'' indicates to KoNViD-1k \cite{KoNViD-1k}, ``VQC'' indicates to LIVE-VQC \cite{LIVE-VQC}, ``Qua'' indicates to LIVE-Qualcomm \cite{LIVE-Qualcomm}, and ``CVD'' indicates to CVD2014 \cite{CVD2014}.}\label{Mixed-Dataset}
\centering
\fontsize{9.2}{9.2}\selectfont
\setlength{\tabcolsep}{0.5mm}{
\begin{tabular}{cclclclclcl}
\toprule[1.1pt]
\multirow{3}{*}{Training set}               & \multicolumn{10}{c}{Testing set}                                                                                                                                             \\ \cmidrule(lr){2-11}
                                         & \multicolumn{2}{c}{CVD} & \multicolumn{2}{c}{KoN} & \multicolumn{2}{c}{Qua} & \multicolumn{2}{c}{VQC} & \multicolumn{2}{c}{Weighted average} \\
                                         & \multicolumn{2}{c}{MDTVSFA \ HVS-5M} & \multicolumn{2}{c}{MDTVSFA \ HVS-5M}   & \multicolumn{2}{c}{MDTVSFA \ HVS-5M}       & \multicolumn{2}{c}{MDTVSFA \ HVS-5M}  & \multicolumn{2}{c}{MDTVSFA \ HVS-5M}              \\ \midrule
KoN+CVD                        & \multicolumn{2}{c}{0.8552 \quad \ \textbf{0.8948}}  & \multicolumn{2}{c}{0.7742 \quad \ \textbf{0.8380}}    & \multicolumn{2}{c}{0.6484 \quad \ \textbf{0.6592}}        & \multicolumn{2}{c}{0.6933 \quad \ \textbf{0.7684}}    & \multicolumn{2}{c}{0.7498 \quad \ \textbf{0.8090}}               \\
KoN+VQC                  & \multicolumn{2}{c}{0.6325 \quad \ \textbf{0.7990} }  & \multicolumn{2}{c}{0.7974 \quad \ \textbf{0.8278}}    & \multicolumn{2}{c}{0.6995 \quad \ \textbf{0.7358}}        & \multicolumn{2}{c}{0.7461 \quad \ \textbf{0.8515}}   & \multicolumn{2}{c}{0.7575 \quad \ \textbf{0.8224}}               \\
KoN+Qua                   & \multicolumn{2}{c}{0.6933 \quad \ \textbf{0.7264}}  & \multicolumn{2}{c}{0.7835 \quad \ \textbf{0.8332}}    & \multicolumn{2}{c}{0.8169 \quad \ \textbf{0.8320}}         & \multicolumn{2}{c}{0.6968 \quad \ \textbf{0.7645}}    & \multicolumn{2}{c}{0.7544 \quad \ \textbf{0.8038}}               \\
Qua+CVD                     & \multicolumn{2}{c}{0.8636 \quad \ \textbf{0.8835}}  & \multicolumn{2}{c}{0.6691 \quad \ \textbf{0.7068}}    & \multicolumn{2}{c}{0.7882 \quad \ \textbf{0.8160}}         & \multicolumn{2}{c}{0.6153 \quad \ \textbf{0.7248}}   & \multicolumn{2}{c}{0.6865 \quad \ \textbf{0.7403}}               \\
Qua+VQC                   & \multicolumn{2}{c}{0.5849 \quad \ \textbf{0.7168}}  & \multicolumn{2}{c}{0.6843 \quad \ \textbf{0.7421}}    & \multicolumn{2}{c}{0.8010 \quad \ \textbf{0.8234}}        & \multicolumn{2}{c}{0.7434 \quad \ \textbf{0.8417}}   & \multicolumn{2}{c}{0.7003 \quad \ \textbf{0.7732}}               \\
VQC+CVD                & \multicolumn{2}{c}{0.8375 \quad \ \textbf{0.8757}}  & \multicolumn{2}{c}{0.7378 \quad \ \textbf{0.7582}}    & \multicolumn{2}{c}{0.6796 \quad \ \textbf{0.7040}}         & \multicolumn{2}{c}{0.7277 \quad \ \textbf{0.8398}}   & \multicolumn{2}{c}{0.7402 \quad \ \textbf{0.7869}}               \\
KoN+CVD+Qua        & \multicolumn{2}{c}{0.8412 \quad \ \textbf{0.8867}}  & \multicolumn{2}{c}{0.7659 \quad \ \textbf{0.8302}}    & \multicolumn{2}{c}{0.8158 \quad \ \textbf{0.8517}}        & \multicolumn{2}{c}{0.6851 \quad \ \textbf{0.7785}}   & \multicolumn{2}{c}{0.7572 \quad \ \textbf{0.8246}}               \\
KoN+VQC+Qua         & \multicolumn{2}{c}{0.6423 \quad \ \textbf{0.7489}}  & \multicolumn{2}{c}{0.7906 \quad \ \textbf{0.8167}}    & \multicolumn{2}{c}{0.8003 \quad \ \textbf{0.8443}}        & \multicolumn{2}{c}{0.7475 \quad \ \textbf{0.8352}}    & \multicolumn{2}{c}{0.7647 \quad \ \textbf{0.8191}}               \\
KoN+VQC+CVD              & \multicolumn{2}{c}{0.8303 \quad \ \textbf{0.8836}}  & \multicolumn{2}{c}{0.7859 \quad \ \textbf{0.8206}}    & \multicolumn{2}{c}{0.7013 \quad \ \textbf{0.7376}}        & \multicolumn{2}{c}{0.7443 \quad \ \textbf{0.8397}}   & \multicolumn{2}{c}{0.7718 \quad \ \textbf{0.8245}}               \\
Qua+VQC+CVD          & \multicolumn{2}{c}{0.8365 \quad \ \textbf{0.8899}}  & \multicolumn{2}{c}{0.7153 \quad \ \textbf{0.7454}}    & \multicolumn{2}{c}{0.8118 \quad \ \textbf{0.8363}}         & \multicolumn{2}{c}{0.7212 \quad \ \textbf{0.8460}}    & \multicolumn{2}{c}{0.7385 \quad \ \textbf{0.7955}}               \\
KoN+VQC+Qua+CVD & \multicolumn{2}{c}{0.8292 \quad \ \textbf{0.8976}}  & \multicolumn{2}{c}{0.7793 \quad \ \textbf{0.8178}}    & \multicolumn{2}{c}{0.8045 \quad \ \textbf{0.8462}}        & \multicolumn{2}{c}{0.7352 \quad \ \textbf{0.8474}}         & \multicolumn{2}{c}{0.7753 \quad \ \textbf{0.8366}}               \\ \bottomrule[1.1pt]
\end{tabular}}
\end{table*}

\begin{table*}[]
\caption{Performance evaluations of ablation studies.}\label{Ablation study}
\centering
\fontsize{9.2}{9.2}\selectfont
\setlength{\tabcolsep}{0.75mm}{

\begin{tabular}{cccccccccccccc}
\toprule[1.1pt]
\multirow{2}{*}{Model}                   & CVD2014                &  & LIVE-Qualcomm   &  & KoNViD-1k  &  & LIVE-VQC    &  & YouTube-UGC  &  & LSVQ-test &  & LSVQ-test-1080p\\ \cmidrule(lr){2-2} \cmidrule(lr){4-4} \cmidrule(lr){6-6} \cmidrule(lr){8-8}  \cmidrule(lr){10-10} \cmidrule(lr){12-12} \cmidrule(lr){14-14}
                                & SRCC \ \   PLCC     &  & SRCC  \ \  PLCC    &  & SRCC \ \   PLCC    &  & SRCC \ \   PLCC      &  & SRCC  \ \  PLCC &  & SRCC  \ \  PLCC &  & SRCC  \ \  PLCC \\ \midrule
Baseline                              & 0.8729 \ \ \textbf{0.8853}  &  & 0.8090   \ \  \textbf{0.8210} &  &  \textbf{0.8530} \ \  \textbf{0.8562}  &  & \textbf{0.8441}  \ \ \textbf{0.8422}  &  &   \textbf{0.8520}   \ \  \textbf{0.8451}  &  &      \textbf{0.8785}   \ \  \textbf{0.8723} &  &      \textbf{0.7977}   \ \ \textbf{0.8172}       \\
Variant I                  & \textbf{0.8795} \ \  0.8799 &  & \textbf{0.8122}  \ \   0.8189 &  &   0.8361 \ \  0.8372  &  & 0.8334 \ \  0.8284  &  &    0.8471   \ \  0.8389  &  &      0.8767 \ \   0.8715  &  & 0.7960  \ \   0.8146  \\

Variant II                           & 0.8619 \ \  0.8627  &  & 0.7729   \ \  0.7789 &  &   0.8096  \ \  0.8012 &  & 0.8139 \ \ 0.8075  &  & 0.8208   \ \  0.8139   &  &      0.8342   \ \  0.8415  &  &      0.7569   \ \  0.7805       \\
Variant III                       & 0.8509 \ \ 0.8523  &  & 0.7911 \ \   0.8081 &  &      0.8379 \ \  0.8371  &  & 0.8323 \ \ 0.8318  &   &  0.8422   \ \ 0.8365 &  &      0.8544   \ \  0.8588  &  &      0.7573   \ \  0.7809    \\

Variant IV                   & 0.8654 \ \ 0.8705 &  & 0.7897  \ \   0.8146 &  &     0.8340 \ \  0.8356 &  & 0.7685 \ \  0.7873   &   & 0.8355   \ \ 0.8352  &  &      0.8479  \ \  0.8538  &  &      0.7542   \ \ 0.7902    \\
Variant V                   & 0.8724 \ \ 0.8753 &  & 0.8072  \ \   0.8124 &  & 0.7314 \ \  0.7273 &  & 0.7989 \ \  0.7750  &   &    0.8136   \ \  0.7825  &  &      0.7940   \ \  0.7732   &  &      0.7556   \ \  0.7585    \\

Variant VI   &     0.8695  \ \        0.8756                         &  &0.8003  \ \        0.8197                   &  &       0.8047  \ \        0.7994  &  &    0.8138  \ \        0.8171 &          &  0.8055  \ \        0.8040    &  &      0.8528   \ \  0.8460  &  &      0.7589   \ \  0.7795     \\
\bottomrule[1.1pt]
\end{tabular}}
\end{table*}

\subsection{Performance Evaluations on Individual Datasets} \label{individual datasets}
In this part, the performance is evaluated on individual datasets, wherein the training set, the validation set, and the testing set come from the same dataset. The results are given in Table \ref{Individual Dataset}, and the following observations can be made.

\begin{itemize}
\item Our proposed HVS-5M achieves SOTA performance on five different datasets. Compared with the existing best-performed BVQA-2022 \cite{Li2022}, the average improvement is 1.5\% and 1.5\% for SRCC and PLCC on six datasets.


\item The improvement over BVQA-2022 is especially significant on the largest LSVQ dataset, which is 3.3\% and 2.8\% for SRCC and PLCC. By contrast, HVS-5M is less effective on the smallest LIVE-Qualcomm dataset. The reason is that its sample quantity may not be enough to activate the potential of HVS-5M with large architecture.

\item The performance of IQA is generally worse than that of VQA, since IQA only operates on spatial domain, and cannot perceive the motion information in temporal domain.
    Such results verify the necessity of the  domain-fusion design paradigm in HVS-5M.
\end{itemize}

\subsection{Performance Evaluations on Categorical Subsets} \label{categorical subsets}

In this part, the performance is evaluated on categorical subsets generated from KoNViD-1k \cite{KoNViD-1k}, LIVE-VQC \cite{LIVE-VQC}, and YouTube-UGC \cite{YouTube-UGC}.
One subset consists of videos with one same attribute, and can emphatically reflect the VQA performance in specific aspect.
According to \cite{VIDEVAL}, three types of categorical subsets are considered as follows.

\begin{itemize}
\item Resolution is an important factor affecting the quality of a video. To further investigate the performance of VQA methods at different resolutions, we select 402 1080p videos, 564 720p videos, and 607 below 480p videos as three subsets. The results are reported in Table \ref{resolution}.

\item Content can also affect the quality of video, which relies on personal preference of human beings. Correspondingly, we select 134 Screen Contents, 70 Animations, and 180 Gamings as three subsets. The results are reported in Table \ref{content}.

\item Evaluation of high quality and low quality videos is an important benchmark for VQA methods. We divide all videos into 1469 low quality videos and 1458 high quality ones by
    threshold 3.5537. The results are reported in Table \ref{quality}.
\end{itemize}

From the above three tables, it can be observed that our proposed HVS-5M achieves the best performance in almost all categorical subsets.
In particular, in the case of Gaming subsets, HVS-5M outperforms BVQA-2022 \cite{Li2022} by 22.0\% and 11.1\% on SRCC and PLCC.
The only exception is the case of subset of high quality video, where HVS-5M is slightly inferior to BVQA-2022.



\begin{table}[]
\caption{Performance evaluations of different models.}\label{Saliency threshold}
\centering
\fontsize{9.2}{9.2}\selectfont
\begin{tabular}{cclclcl}
\toprule[1.1pt]
\multirow{2}{*}{Different models}    & \multicolumn{2}{c}{CVD2014}         & \multicolumn{2}{c}{} & \multicolumn{2}{c}{LIVE-Qualcomm}   \\ \cmidrule(lr){2-3} \cmidrule(lr){6-7}
    & \multicolumn{2}{c}{SRCC \quad  PLCC}     & \multicolumn{2}{c}{} & \multicolumn{2}{c}{SRCC  \quad PLCC}     \\  \midrule
$h=30$  & \multicolumn{2}{c}{0.8778 \quad  0.8851} & \multicolumn{2}{c}{} & \multicolumn{2}{c}{0.8110 \quad  0.8207}  \\
$h=50$  & \multicolumn{2}{c}{0.8820 \quad   0.8908}  & \multicolumn{2}{c}{} & \multicolumn{2}{c}{\textbf{0.8163}  \quad \textbf{0.8226}} \\
$h=60$  & \multicolumn{2}{c}{\textbf{0.8854} \quad  \textbf{0.8934}} & \multicolumn{2}{c}{} & \multicolumn{2}{c}{0.8149 \quad  0.8201} \\
$h=80$  & \multicolumn{2}{c}{0.8801 \quad  0.8903} & \multicolumn{2}{c}{} & \multicolumn{2}{c}{0.8067  \quad 0.8148} \\
$h=100$ (baseline) & \multicolumn{2}{c}{0.8729 \quad  0.8853} & \multicolumn{2}{c}{} & \multicolumn{2}{c}{0.8090  \quad 0.8210}   \\
Variant I  & \multicolumn{2}{c}{0.8795 \quad  0.8799} & \multicolumn{2}{c}{} & \multicolumn{2}{c}{0.8122  \quad 0.8189} \\  \bottomrule[1.1pt]
\end{tabular}
\end{table}


 \subsection{Performance Evaluations on Cross Datasets} \label{cross datasets}
In this part, the performance is evaluated on cross datasets. Since different video datasets have great differences in duration, content, and resolution, evaluations on cross datasets can verify the generalization ability and robustness of VQA methods.
Following the settings in \cite{Chen2021} and \cite{LSVQ}, we conduct two experiments on cross datasets, and the results are given in Table \ref{Cross-Dataset1} and Table \ref{Cross-Dataset2}, respectively. The following observations can be made.


\begin{itemize}
\item Table \ref{Cross-Dataset1} shows the results of training models on four relatively smaller datasets. It can be observed that our proposed HVS-5M outperforms existing methods in most cases. Especially, in the case of being trained on LIVE-VQC and tested on LIVE-Qualcomm, HVS-5M outperforms CNN-TLVQM \cite{CNN-TLVQM} by 13.3\% and 8.1\% on SRCC and PLCC, respectively, demonstrating that our method has stronger generalization ability on different datasets.

\item Table \ref{Cross-Dataset2} shows the results of training models on the largest LSVQ dataset. It can be observed that in the case of being tested on KoNViD-1k, the proposed HVS-5M outperforms BVQA-2022 \cite{Li2022} by 2.1\% and 3.0\% on SRCC and PLCC, while in the case of being tested on LIVE-VQC, the proposed HVS-5M obtains comparable performance as BVQA-2022.

\end{itemize}

\begin{figure}[t]\centering

\setlength{\belowcaptionskip}{-0.18cm}
\centerline{\includegraphics[width=3.8in]{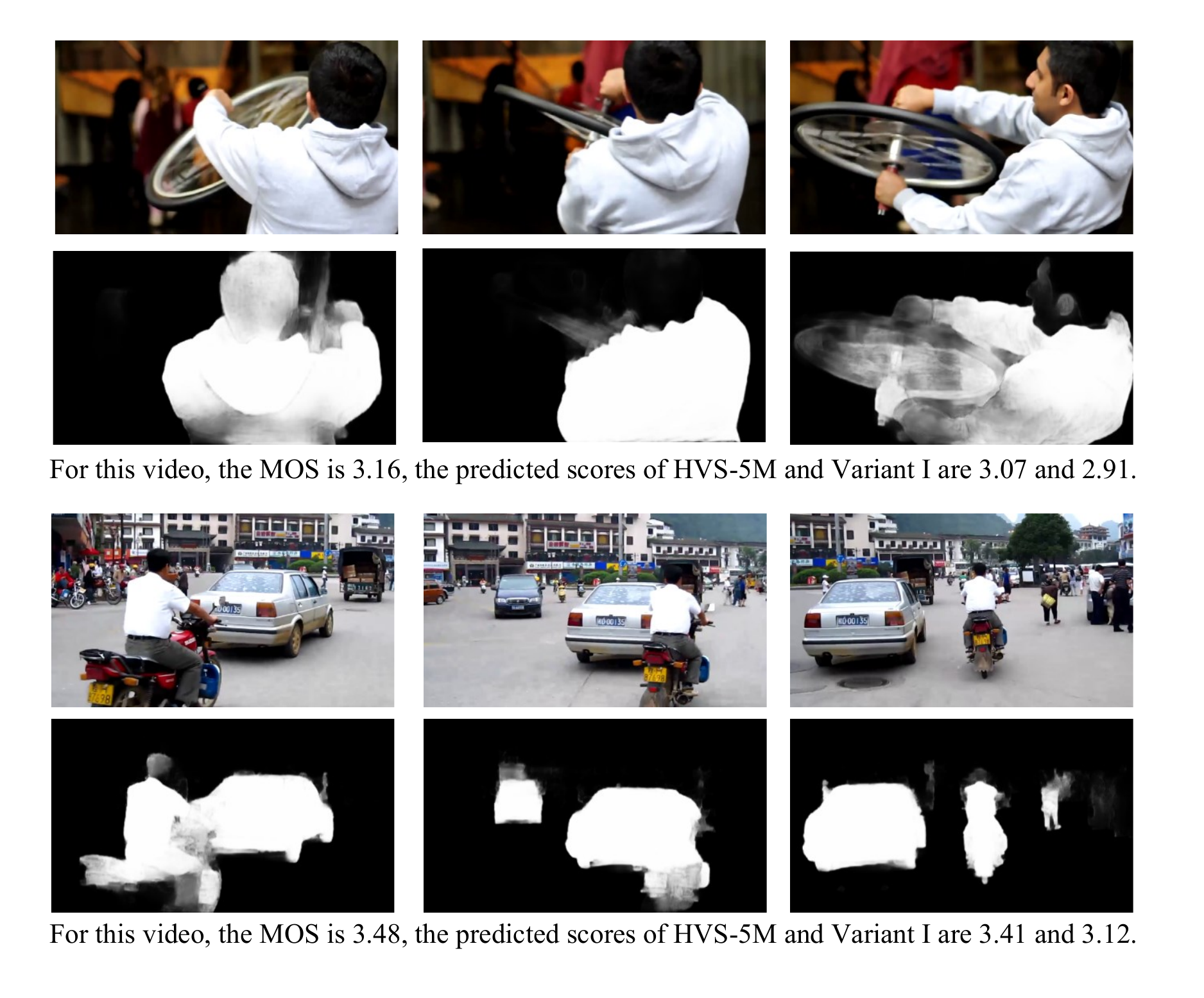}}
\setlength{\abovecaptionskip}{-0.15cm}

\caption{The video frame and its corresponding saliency map.} \label{Variant saliency}
\end{figure}

 \subsection{Performance Evaluations on Mixed Datasets} \label{mixed datasets}
In this part, the performance is evaluated on mixed datasets. Nowadays, most of the methods based on deep learning are data-driven, and their performance relies on the quantity of datasets. Therefore, it is beneficial to utilize multiple datasets for mixed training, and thus the evaluations on mixed datasets are of practical value.
The results are given in Table \ref{Mixed-Dataset}.


\begin{itemize}
\item Compared with MDTVSFA \cite{MDTVSFA}, the SRCC of our proposed HVS-5M shows significant improvement in all cases. In particular, in the case of being trained on the mixed dataset of KoNViD-1k and LIVE-VQC, and tested on CVD2014, HVS-5M achieves
    26.3\% improvement over MDTVSFA on SRCC. 

\item However, increasing the quantity of training data may not necessarily bring performance improvement for both our HVS-5M and MDTVSFA.
    For example, regarding the mixture of KoNViD-1k and LIVE-VQC as the baseline training set. The performance of HVS-5M tested on KoNViD-1k drops from 0.8278 to 0.8167 when adding LIVE-Qualcomm into the training set. Similar results can be observed from MDTVSFA. The reason is that the added training set LIVE-Qualcomm and the testing set KoNViD-1k have great difference in resolution and time duration. This indicates that blindly increasing the quantity of training samples may not be the best choice. Instead, we should consider the distribution of the training set and the testing set.
\end{itemize}




\subsection{Ablation studies} \label{ablation studies}



In this part, ablation studies are conducted on the proposed framework. Specifically, HVS-5M is taken as baseline, and the impact of different modules are investigated by Variants I, II, III, IV, and V, respectively, and the impact of CNN architectures are investigated by Variant VI.




\begin{enumerate}[]
\item  \textit{Variant I.} The visual saliency module is disabled.
\item  \textit{Variant II.} The content-dependency module is disabled.
\item  \textit{Variant III.} The edge masking module is disabled.
\item  \textit{Variant IV.} The motion perception module is disabled.
\item  \textit{Variant V.} The fully-connected network is utilized to obtain the quality score, rather than TempHyst in the temporal hysteresis module.
\item  \textit{Variant VI.} ConvNeXt is replaced with ResNet-50 in the content-dependency and the edge masking module.

\end{enumerate}

The experimental results are presented in Table \ref{Ablation study}, and the following observations can be made from these results.


\begin{itemize}
\item The proposed HVS-5M outperforms the variants in most cases, indicating that HVS-5M can play its potential to full play when the five characteristics in the revisited HVS work in the collaborative manner.



\item Disabling the visual saliency module in Variant I brings performance degradation on four datasets, except LIVE-Qualcomm and CVD2014.
    Note that $h$ is set to 100 according to the debug experiments on KoNViD-1k, and HVS-5M with such setting has achieved SOTA performance on five different datasets.
    The setting of $h$ depends on the characteristics of datasets, and further debugging $h$ can bring greater improvement on specific dataset.
    As shown in Table \ref{Saliency threshold}, setting $h$ to 50 and 60 are better options on LIVE-Qualcomm and CVD2014, respectively.
    Besides, from Fig. \ref{Variant saliency}, it can be observed that our HVS-5M indeed focuses on the significant regions within video.

\item Disabling the content-dependency module in Varint II and disabling the edge masking module in Varint III lead to average performance degradation of 4.2\% and 2.5\% for SRCC on six datesets, respectively. Specifically, on CVD2014, the performance degradation by Variant III is even more significant than that by Variant II, indicating that the edge features may be more essential than the content features under certain circumstances.


\item Replacing TempHyst with fully-connected network in the temporal hysteresis module in Variant V brings the most significant performance degradation among all variants.
    Specifically, the degradations are 16.6\% and 17.7\% for SRCC and PLCC on KoNViD-1k, indicating that the memory mechanism in such module plays an important role in HVS-5M.

\item Replacing ResNet-50 with ConvNeXt in the content-dependency and the edge masking module in Variant VI also leads to significant performance degradation of 3.5\% and 2.8\% for SRCC and PLCC averaging on six datasets.
    It can be inferred that our proposed HVS-5M framework is flexible, and its performance can be further boosted with more advanced network architectures.

\end{itemize}

\section{Conclusions} \label{conclusion}
In this paper, we propose a new NRVQA framework called HVS-5M. The foundation of HVS-5M is the revisited HVS, wherein five well-connected characteristics are utilized to model the function of sensory organ in a relatively simple and comprehensive manner.
HVS-5M follows the domain-fusion design paradigm, and simultaneously extracts the frame-level spatial features and the video-level temporal features, which are integrated to obtain the final quality score.
Extensive experiments have been conducted to evaluate the performance of HVS-5M, and the following conclusions can be drawn.
\begin{enumerate}[]
\item Evaluations on individual dataset show that HVS-5M achieves the SOTA performance on various mainstream datasets.

\item Evaluations on categorical subsets, cross datasets, and mixed datasets show that HVS-5M has strong generalization ability and robustness.

\item Ablation studies verify the contributions of five characteristics in HVS-5M, and the effectiveness of ConvNeXt.
\end{enumerate}

To the best of our knowledge, in the area of VQA, this paper makes the first attempt by introducing the characteristics of edge masking, and a new scheme to apply visual saliency.
The proposed framework represents a promising direction for objective VQA. Further investigations may include the following aspects.
Firstly, more characteristics of HVS can be explored to better simulate the function of sensory organ.
Secondly, there is still improvement room for the application of visual saliency.
Thirdly, to bring the potential of learning ability to full play, it is desired to train the model in an end-to-end manner.

\end{document}